\documentclass[amsmath,amssymb,amsbsy,prx,nofootinbib,twocolumn,tightenlines,superscriptaddress,floatfix]{revtex4-2}
\usepackage{hyperref}
\usepackage{graphicx}
\usepackage{amsfonts,amssymb,amsmath,amsbsy}
\usepackage[caption=false, justification=centerlast]{subfig}
\usepackage{color}
\usepackage{tikz}
\usepackage[normalem]{ulem}
\usetikzlibrary{arrows.meta,positioning,calc,decorations.pathmorphing}
\tikzset{
  flow/.style={flowblue, line width=1.15pt, -{Stealth[length=3.2mm,width=2.2mm]}},
  dim/.style={black, line width=1.1pt, {Stealth[length=2.8mm,width=2.0mm]}-{Stealth[length=2.8mm,width=2.0mm]}},
  lab/.style={font=\large},
  epslab/.style={font=\Large}
}
\definecolor{blobdark}{RGB}{153,187,242}
\definecolor{bloblight}{RGB}{204,219,249}
\definecolor{flowblue}{RGB}{28,113,225}

\graphicspath{{./}}
\newcommand{\bv}{\boldsymbol{v}}

\newcommand{\bx}{\boldsymbol{x}}

\newcommand{\bX}{\boldsymbol{X}}

\newcommand{\dd}{\mathrm{d}}

\newcommand{\beps}{\bar{\varepsilon}}

\begin{document}
	
	\title{Contrasting Eulerian and Lagrangian organization of turbulent dissipation}
	
	\newcommand{\inphyni}{Universit\'e C\^ote d'Azur, CNRS, Inria, INPHYNI, 06100 Nice, France}
	\newcommand{\icts}{International Centre for Theoretical Sciences, Tata Institute of Fundamental Research, Bengaluru 560089, India}
	\newcommand{\iitk}{Department of Mechanical Engineering, Indian Institute of Technology Kanpur, Kanpur 208016, India}
	\author{J\'er\'emie Bec$^\clubsuit$}
	\email{jeremie.bec@univ-cotedazur.fr}
	\affiliation{\inphyni}
	\author{Ritwik Mukherjee$^\clubsuit$}
	\email{ritwik.mukherjee@icts.res.in}
	\affiliation{\icts}
	\author{Siddhartha Mukherjee}
	\email{smukherjee@iitk.ac.in}
	\affiliation{\iitk}
	\author{Samriddhi Sankar Ray}
	\email{samriddhisankarray@gmail.com}
	\affiliation{\icts}
	
	\date{\today}
	
\begin{abstract}
The dissipation of turbulent kinetic energy is conventionally attributed to a scale-invariant spatial multiplicative cascade, with Lagrangian statistics inferred through bridge relations. Direct numerical simulations with tracers reveal instead that Eulerian coarse-grained dissipation has a squared-logarithmic cascade depth and scale-dependent multipliers, whereas time-averaged dissipation along trajectories displays multifractal scaling and an approximately scale-invariant temporal cascade. The classical bridge fails at the level of individual trajectories. Eulerian multipliers nevertheless possess a multifractal organization whose accumulation accounts for the squared-logarithmic spatial statistics. This contrast motivates a material-inheritance picture in which Eulerian dissipation statistics reflect the spatial assembly, deformation, and mixing of distinct Lagrangian histories. The results therefore distinguish the temporal cascade organization revealed along fluid trajectories from the non-stationary spatial refinement observed in the Eulerian frame.
\end{abstract}
	
\maketitle

\def\thefootnote{$\clubsuit$}\footnotetext{These authors contributed equally to this work}\def\thefootnote{\arabic{footnote}}

%%%%%%%%%%%%%%%%%%%%%%%%%%%%%%%%%%%%%%
\section{Introduction}
\label{sec:intro}
Turbulence thrives on paradox. Time and again, elegant theoretical frameworks built on universal laws or scaling principles have seemed to capture the essence of turbulent motion, only to be challenged by new experiments or simulations revealing subtle yet profound deviations from the expected behavior. Kolmogorov’s 1941 theory provides perhaps its most celebrated example: assuming that inertial-range statistics depend only on the mean energy-transfer rate and the observation scale, it predicts universal self-similar scaling, with velocity structure-function exponents $\zeta_p=p/3$~\cite{Frisch_1995}. This phenomenology has extraordinary explanatory power, yet experiments and numerical simulations consistently reveal intense rare fluctuations, increasingly non-Gaussian statistics toward small scales, and scaling exponents that depart from the dimensional prediction~\cite{yeung2015extreme,buaria2019extreme}. Intermittency is therefore not merely a small correction to K41, but signals that the energy transfer becomes progressively more heterogeneous along the cascade~\cite{AlexakisBiferaleReview}. The refined similarity hypotheses of Kolmogorov and Obukhov incorporate this heterogeneity through the locally coarse-grained dissipation $\varepsilon_\ell$, relating velocity increments to the fluctuating local transfer rate through $\delta_\ell v\sim(\varepsilon_\ell\ell)^{1/3}$~\cite{kolmogorov1962refinement,obukhov1962some}. Anomalous velocity scaling is thereby connected to the statistics of $\varepsilon_\ell$, placing the Eulerian dissipation field at the center of intermittency phenomenology.

This connection has encouraged two notions that are closely related but not equivalent. Multifractal scaling is first of all a statement about moments: the normalized moments of the coarse-grained dissipation behave as $\langle(\varepsilon_\ell/\beps)^q\rangle\propto(\ell/L)^{\chi_q}$, with a nonlinear dependence of $\chi_q$ on $q$ reflecting intermittency~\cite{Frisch_1995,meneveau1987multifractal,meneveau1991multifractal}. A scale-stationary multiplicative cascade makes a stronger assertion about the process across scales: passing from one scale to another amounts statistically to multiplication by a random factor whose law depends only on the scale ratio, so that equal intervals of logarithmic scale are statistically equivalent. In its log-divisible form, the logarithmic increments accumulated over disjoint scale intervals are also independent. Such a process naturally produces multifractal moment scaling, but the converse need not hold: observing power-law moments over a finite range does not, by itself, establish an underlying scale-stationary, log-divisible process. This distinction will be important when comparing the spatial and temporal organizations of dissipation fluctuations.

The multifractal description of Eulerian dissipation has a substantial empirical basis. Early experiments, relying largely on one-dimensional surrogates reconstructed from hot-wire velocity measurements, found nonlinear moment exponents and consistent multifractal spectra across different flows and Reynolds numbers~\cite{meneveau1987multifractal,meneveau1991multifractal}. Direct numerical simulations later made it possible to evaluate the full tensorial dissipation and to perform genuinely three-dimensional local averages. These studies provided direct support for refined similarity~\cite{iyer2015refined} and reported anomalous scaling of the moments of gradients~\cite{Schumacher2007,donzis2008dissipation,Dey2026}  
and locally averaged dissipation at Reynolds numbers up to $R_\lambda\simeq1300$~\cite{iyer2019scaling,buaria2022intermittency}. In particular, the second-order intermittency exponent was found to approach the commonly reported value $\mu\simeq0.23$ at the highest Reynolds numbers. Multifractal moment scaling of the Eulerian dissipation field is therefore supported by several generations of measurements, although the measured observable, coarse-graining geometry, and accessible scale range have varied among studies.

\begin{figure*}[t!]
\includegraphics[width=1\linewidth]{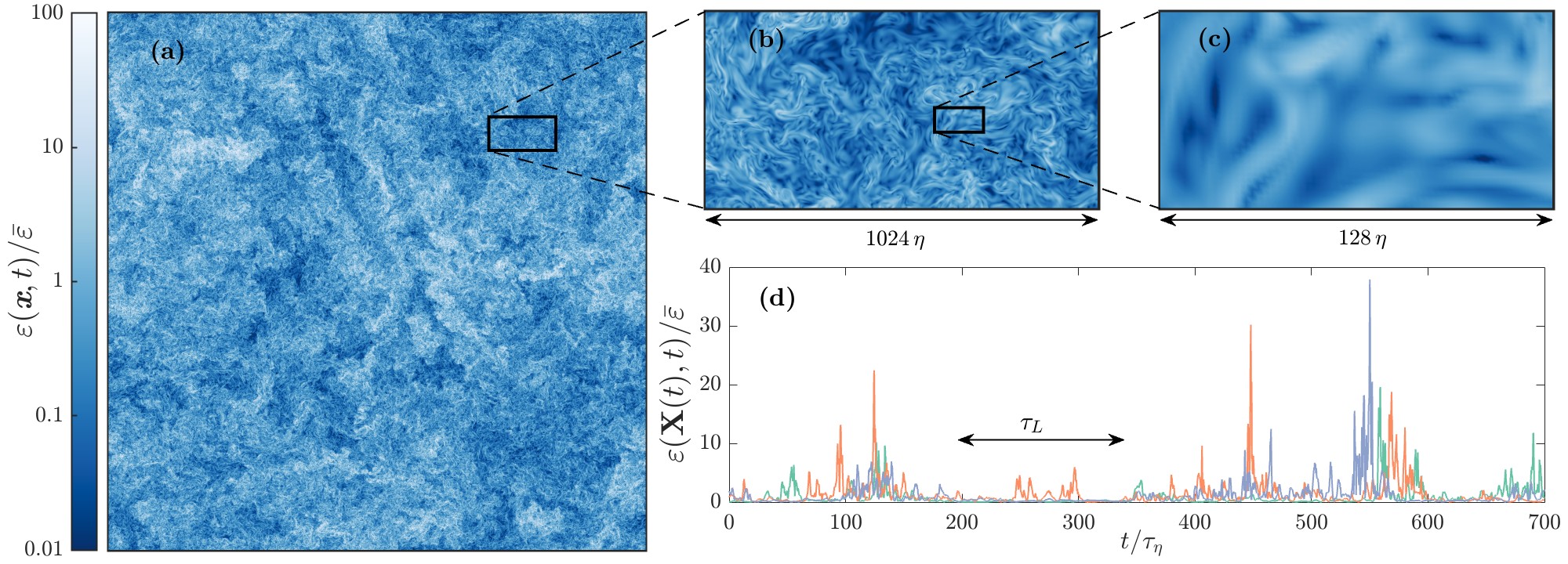}
\caption{(a) Snapshots of the local Eulerian energy dissipation field $\varepsilon(\boldsymbol{x},t) = (\nu/2)\|\nabla\bv+\!\nabla\bv^{\!\top}\!\|^2$ in a two-dimensional slice of the simulation at $R_\lambda=730$, shown at increasing magnification in regions of linear size (b) $1024\,\eta$ and (c) $128\,\eta$. Intense dissipation events appear highly localized and filamentary in space.  (d) Dissipation sampled along Lagrangian tracer trajectories, $\tilde{\varepsilon}(t)=\varepsilon(\boldsymbol{X}(t),t)$, shown for the same flow. Individual tracers experience strongly intermittent temporal fluctuations.}
\label{fig:snapshot_dissip}
\end{figure*}
Evidence bearing on the stronger multiplicative-cascade interpretation has come from additional diagnostics. Hot-wire measurements of surrogate-dissipation multipliers found probability distributions that were approximately independent of the cascade level, supporting the idea that successive intervals of logarithmic scale are statistically equivalent~\cite{chhabra1992scale}. Subsequent analyses nevertheless detected correlations between multipliers at successive levels~\cite{sreenivasan1995turbulent}, showing that scale-invariant multiplier distributions alone do not establish a log-divisible process with independent increments. A complementary line of investigation has examined correlations of the logarithm of dissipation, whose logarithmic dependence on spatial separation is another natural signature of multiplicative cascades~\cite{arneodo1998analysis}. Recent analysis of three-dimensional DNS by Ruffenach and Chevillard reported spatial and Eulerian temporal statistics compatible with such log-correlated fields and developed their representation in terms of Gaussian multiplicative chaos~\cite{ruffenach2026spatiotemporal}. Together, these results motivate examining moment scaling, multipliers, and logarithmic correlations as related but logically distinct signatures of a cascade.

The Lagrangian viewpoint, which follows individual fluid particles rather than
observing the flow at fixed spatial positions,
provides a complementary temporal description of intermittency~\cite{ToschiBodenschatzARFM}.
Particle-tracking experiments and direct numerical simulations have shown that
Lagrangian velocity increments and accelerations undergo extreme fluctuations,
often displaying stronger intermittency than their Eulerian
counterparts~\cite{sawford2011kolmogorov,biferale2005lagrangian,sawford2015direct,arneodo2008universal,bentkamp2019persistent}.
Dissipation sampled along trajectories has received less attention, but the
study of Huang and Schmitt~\cite{huang2014lagrangian} provided several
indications of a temporal cascade. Using DNS at $R_\lambda=400$, they found
approximately log-normal statistics for both the instantaneous and
time-averaged Lagrangian dissipation. The moments of the time-averaged
dissipation exhibited power-law scaling over $10<\tau/\tau_\eta<100$, with
exponents consistent with those inferred from Lagrangian velocity statistics.
Moreover, both the autocorrelation of $\ln[\tilde{\varepsilon}(t)/\beps]$ and
the variance of $\ln(\tilde{\varepsilon}_\tau/\beps)$ followed logarithmic laws
characterized by a common intermittency coefficient $\mu\simeq0.30$. The
concurrence of these signatures led to the proposal that dissipation along
fluid trajectories can be described by a multiplicative cascade evolving in
time.

Despite these parallel developments, the relation between the spatial and temporal organizations of dissipation remains unclear. Eulerian and Lagrangian intermittency are traditionally connected through \emph{bridge relations}, which associate a spatial scale with a characteristic eddy-turnover time and thereby infer temporal scaling from spatial statistics~\cite{borgas1993multifractal,schmitt2006linking,chevillard2012phenomenological,lalescu2018tracer}. Although this approach has been successful for velocity fluctuations, its extension to dissipation is less direct. The Eulerian quantity $\varepsilon_\ell(\boldsymbol{x},t)$ averages the dissipation over a spatial region at a fixed time, whereas its Lagrangian counterpart averages the values accumulated by a moving particle through an evolving flow. A correspondence between their coarse-graining scales, and still more between their multiplier statistics, does not therefore follow from dimensional arguments alone. As illustrated in Fig.~\ref{fig:snapshot_dissip}, the same dissipation field appears as highly localized spatial structures in the Eulerian frame~\cite{yeung2015extreme,mukherjee2024turbulent,vassilicos2023beyond,mukherjee2026} and as intermittent temporal bursts along trajectories. Whether these two manifestations are governed by a common scale-stationary cascade mechanism is the central question addressed here.

To address this question, high-resolution direct numerical simulations of homogeneous isotropic turbulence seeded with Lagrangian tracers are analyzed. The incompressible Navier--Stokes equations are integrated using a pseudo-spectral method for three datasets, with resolutions up to $4096^3$ and Taylor-scale Reynolds numbers $R_\lambda=315$, $420$, and $730$. The numerical parameters are summarized in Table~\ref{tab:numparam}.
\begin{table}[h!]
\caption{\label{tab:numparam}\emph{DNS parameters.}\/ $\eta=\nu^{3/4}/\beps^{1/4}$ is the Kolmogorov dissipative length scale, with $\beps$ the average dissipation rate, $\lambda=\sqrt{15\nu\,u_{\mathrm{rms}}^2/\beps}$ the Taylor microscale, and $L=u_{\mathrm{rms}}^3/\beps$ the integral length scale, with $u_{\mathrm{rms}}=\sqrt{\langle|\boldsymbol{v}|^2\rangle/3}$.  The corresponding time scales are the Kolmogorov time $\tau_\eta=\sqrt{\nu/\beps}$ and the large-eddy turnover time $\tau_L=u_{\mathrm{rms}}^2/\beps$. The Taylor-scale Reynolds number is $R_\lambda=u_{\mathrm{rms}}\lambda/\nu$.
}
\begin{ruledtabular}
	\begin{tabular}{ccccccc}
		$N^3$       &   $\eta$     & $\lambda$ & $L$   & $\tau_\eta$ & $\tau_L$ & $R_\lambda$\\ \colrule
%		$512^3$   & $0.005$     & $0.17$      & $1.17$ & $0.035$   & $1.98$   & 220\\
		$1024^3$ & $0.0020$   & $0.073$    & $1.53$ & $0.022$   & $1.76$   & $315$\\
		$2048^3$ & $0.0027$   & $0.11$      & $3.03$ & $0.021$   & $2.25$   & $420$\\
		$4096^3$ & $0.00074$ & $0.039$    & $1.86$ & $0.051$   & $9.68 $  & $730$\\
%		$8192^3$ & $0.00051$ & $0.036$    & $1.18$ & $0.006$   & $1.95$   & $1285$
	\end{tabular}
\end{ruledtabular}
\end{table}

The velocity field is denoted by $\bv(\bx,t)$, and the local kinetic-energy dissipation rate is $\varepsilon(\bx,t)=2\nu S_{ij}S_{ij}$, where $S_{ij}=(\partial_jv_i+\partial_iv_j)/2$ and $\nu$ is the kinematic viscosity. Its Eulerian coarse-grained form $\varepsilon_\ell(\bx,t)$ is obtained by averaging $\varepsilon$ over a ball of size $\ell$ centered at $\bx$. Lagrangian tracers satisfy $\dd\bX/\dd t=\bv(\bX,t)$, sample the instantaneous dissipation $\tilde{\varepsilon}(t)=\varepsilon(\bX(t),t)$, and define $\tilde{\varepsilon}_\tau$ by averaging along a trajectory over an interval $\tau$. For uniformly distributed tracers in a statistically stationary, homogeneous, incompressible flow, all these quantities share the mean $\beps=\langle\varepsilon\rangle$.

These observables are examined using the same hierarchy of statistical diagnostics in the two frames. Moment scaling tests for multifractality, while logarithmic cumulants and correlations, together with multiplier statistics, probe the stronger requirements of scale stationarity and log-divisibility. Large-deviation spectra provide a complementary characterization of rare fluctuations and identify the effective cascade depth governing their accumulation across scales~\cite{Frisch_1995,vulpiani2014large}. Rather than assuming an Eulerian--Lagrangian bridge from the outset, this comparison asks which scale variable renders the statistics stationary in each frame and how the resulting cascade organizations are related.

Three principal findings emerge. First, over the resolved range of Reynolds numbers and scales, Eulerian coarse-grained dissipation does not display the organization of a conventional scale-stationary multifractal cascade. Its spatial covariance, logarithmic cumulants, large-deviation statistics, and moment slopes are instead consistently governed by an effective depth proportional to $[\log(\ell/\ell_{\rm c})]^2$. Second, dissipation averaged along Lagrangian trajectories exhibits the standard signatures of a temporal cascade: power-law moment scaling, logarithmic temporal correlations, a large-deviation principle with depth proportional to $\log(\tau_{\rm c}/\tau)$, and approximately scale-stationary multipliers. The associated rate function is nevertheless non-quadratic, revealing a multifractal but non-log-normal temporal process. Third, the classical bridge between spatial and temporal dissipation averages fails as a trajectory-wise identification. Eulerian multipliers are not scale invariant, but their scale dependence has a multifractal organization of its own; the accumulation of these increasingly intermittent refinements accounts for the squared-logarithmic Eulerian depth. This structure motivates a dynamical-inheritance picture in which spatial dissipation patterns result from the deformation and mixing of distinct Lagrangian histories.

The remainder of the paper is organized as follows. Section~\ref{sec:multifractal} introduces the statistical diagnostics used throughout. Sections~\ref{sec:eulerian} and~\ref{sec:lagrangian} present the Eulerian and Lagrangian analyses, respectively. Section~\ref{sec:bridge} revisits bridge relations and proposes a dynamical-inheritance picture, before the conclusions are presented in Section~\ref{sec:conclusion}.

%%%%%%%%%%%%%%%%%%%%%%%%%%%%%%%%%%%%%%
\begin{figure}[b!]
\centering
\resizebox{\linewidth}{!}{\begin{tikzpicture}[
    scale=0.55, transform shape,
    x=1cm,y=1cm,
    cascade node/.style={
        circle,
        draw=blue!70,
        fill=blue!18,
        line width=0.9pt,
        minimum size=8.5mm,
        inner sep=0pt,
        font=\Large
    },
    root node/.style={cascade node,minimum size=11mm},
    branch/.style={-{Stealth[length=2.2mm,width=1.5mm]},draw=gray!70,line width=0.75pt},
    scale label/.style={anchor=west,font=\large},
    multiplier/.style={font=\large,anchor=south east},
    formula box/.style={
        draw=gray!65,
        rounded corners=2mm,
        fill=gray!4,
        inner xsep=6mm,
        inner ysep=3mm
    }
]

% Vertical levels
\def\yroot{7.3}
\def\yone{5.7}
\def\ytwo{3.9}
\def\ythree{2.1}
\def\ydiss{0.35}

% Root and first three generations
\node[root node] (r) at (0,\yroot) {$\varepsilon_0$};

\node[cascade node] (a1) at (-2.4,\yone) {$\varepsilon_1$};
\node[cascade node] (a2) at ( 2.4,\yone) {$\varepsilon_1$};

\foreach \x/\name in {-3.6/b1,-1.2/b2,1.2/b3,3.6/b4}
  \node[cascade node] (\name) at (\x,\ytwo) {$\varepsilon_2$};

\foreach \x/\name in {-4.2/c1,-3.0/c2,-1.8/c3,-0.6/c4,0.6/c5,1.8/c6,3.0/c7,4.2/c8}
  \node[cascade node] (\name) at (\x,\ythree) {$\varepsilon_3$};

% Branches
\draw[branch] (r) -- (a1);
\draw[branch] (r) -- (a2);

\draw[branch] (a1) -- (b1);
\draw[branch] (a1) -- (b2);
\draw[branch] (a2) -- (b3);
\draw[branch] (a2) -- (b4);

\draw[branch] (b1) -- (c1);
\draw[branch] (b1) -- (c2);
\draw[branch] (b2) -- (c3);
\draw[branch] (b2) -- (c4);
\draw[branch] (b3) -- (c5);
\draw[branch] (b3) -- (c6);
\draw[branch] (b4) -- (c7);
\draw[branch] (b4) -- (c8);

% Continuation toward the dissipation range
\foreach \i in {1,...,8}{
  \pgfmathsetmacro{\xc}{-4.8+1.2*\i}
  \draw[branch] (c\i.south west) -- ++(-0.2,-1.0);
  \draw[branch] (c\i.south east) -- ++( 0.2,-1.0);
}

% Multipliers along the left branch
\node[multiplier] at (-1.0,6.6) {$\mathcal{M}_1$};
\node[multiplier] at (-3.1,4.75) {$\mathcal{M}_2$};
\node[multiplier] at (-4.0,2.82) {$\mathcal{M}_3$};

% Dotted indication of unresolved smaller scales
\node[font=\large] at (0,0.18) {$\cdots\quad \cdots\quad \cdots\quad \cdots\quad \cdots\quad \cdots\quad \cdots\quad \cdots\quad \cdots\quad \cdots$};

% Scale annotations
\node[scale label,text=blue!80!black,font=\bfseries\large] at (5.05,\yroot)
  {Integral Scale $(L)$};
\node[scale label] at (5.05,\yone)
  {\textbf{Scale} $\ell_1=L/2$};
\node[scale label] at (5.05,\ytwo)
  {\textbf{Scale} $\ell_2=L/2^2$};
\node[scale label] at (5.05,\ythree)
  {\textbf{Scale} $\ell_3=L/2^3$};
\node[scale label,text=cyan!70!blue,font=\bfseries\large] at (5.05,0.18)
  {Dissipation Range $(\eta)$};

% Cascade relation
%\node[formula box,font=\large] at (0,-1.05)
%  {$\displaystyle
%    \varepsilon_n
%    =\varepsilon_L\,\mathcal{M}_1\mathcal{M}_2\cdots\mathcal{M}_n
%    =\varepsilon_L\prod_{i=1}^{n}\mathcal{M}_i
%  $};

\end{tikzpicture}}
\caption{Schematic diagram of a multiplicative cascade.}
\label{fig:cascade}
\end{figure}
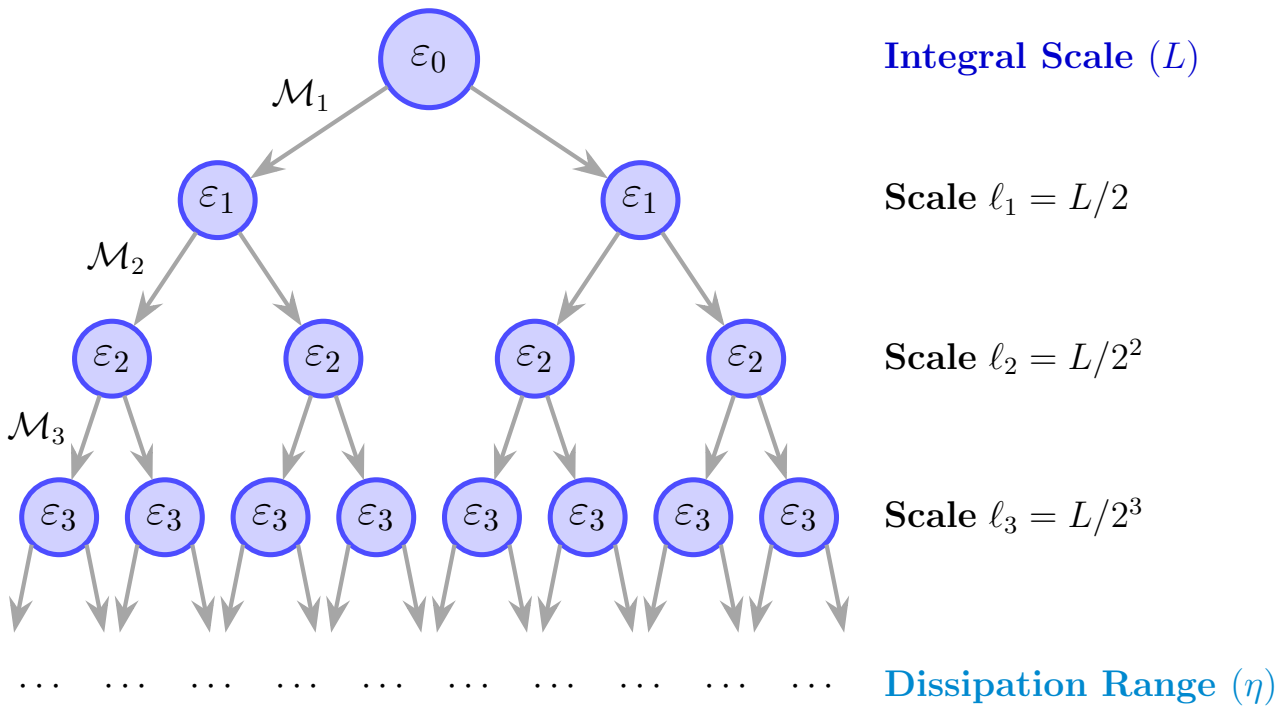
%%%%%%%%%%%%%%%%%%%%%%%%%%%%%%%%%%%%%%

%%%%%%%%%%%%%%%%%%%%%%%%%%%%%%%%%%%%%%
\section{Cascades, large deviations, and multifractal scaling}
\label{sec:multifractal}
Multiplicative cascades provide the standard phenomenological framework for describing intermittency in turbulence. Their statistical structure is naturally formulated in terms of a cascade depth $n$. When this depth is proportional to the logarithmic separation between physical scales, the cascade gives rise to logarithmic cumulants and correlations, together with multifractal power-law scaling. As these properties are examined in the following sections, we briefly summarize their connections.

Following the phenomenology introduced by Kolmogorov and Obukhov~\cite{kolmogorov1962refinement,obukhov1962some} (see also~\cite{Frisch_1995}), a positive quantity, such as the coarse-grained dissipation, is successively refined according to
\begin{equation}
\varepsilon_n = \varepsilon_0\, {\textstyle \prod_{i=1}^{n} \mathcal{M}_i},
\end{equation}
where the random multipliers $\mathcal{M}_i$ describe the transfer of fluctuations between successive cascade levels. The cascade depth $n$ need not be identified \textit{a priori} with a particular physical scale. In the standard scale-invariant construction with a fixed cascade ratio $r_\mathrm{c}>1$, the scale after $n$ refinement steps is $s_n=S/r_\mathrm{c}^n$ where $S$ denotes a reference scale. We then get $n(s)=\log(S/s)/\log r_\mathrm{c}$ and $\varepsilon_s=\varepsilon_{n(s)}$. Figure~\ref{fig:cascade} illustrates the dyadic case $r_\mathrm{c}=2$. More generally, an effective cascade depth $n(s)$ may depend differently on the physical scale, for instance as an arbitrary function of its logarithm.

Taking logarithms transforms the multiplicative cascade into an additive process,
\begin{equation}
Y_n := \log(\varepsilon_n/\varepsilon_0) = {\textstyle \sum_{i=1}^{n} \mathcal{X}_i}, \qquad  \mathcal{X}_i := \log \mathcal{M}_i.
\label{eq:cascade_sum}
\end{equation}
If the logarithmic increments $\mathcal{X}_i$ are independent and stationary, then $Y_n$ has independent and stationary increments in the cascade depth. In a continuous-depth description, it forms a Lévy process whose cumulant generating function takes the form
\begin{equation}
\log \left\langle \mathrm{e}^{q Y_n} \right\rangle = n\,\Psi(q),
\end{equation}
where the per-unit-depth exponent $\Psi$ admits the Lévy--Khintchine representation~\cite{bertoin1996levy}
\begin{equation}
\Psi(q) = \mu\,q + \frac{\sigma^2}{2}\,q^2 + \int_{\mathbb{R}} \left( \mathrm{e}^{qx} - 1 - qx\,\openone_{|x|<1} \right)\rho(\dd x).
\label{eq:levy-khintchine}
\end{equation}
Here $\mu$ is the mean logarithmic increment per unit cascade depth, $\sigma^2$ is the variance of its Gaussian component, and $\rho$ is the L\'evy measure describing jump contributions. The classical log-normal and log-Poisson models correspond respectively to purely Gaussian increments and to jump processes with discrete amplitudes~\cite{she1994universal,dubrulle1994intermittency}.

Provided the corresponding derivatives of $\Psi$ exist, all cumulants of $Y_n$ grow linearly with the cascade depth,
\begin{equation}
\mathcal{K}_m[Y_n] = n\,\Psi^{(m)}(0).
\end{equation}
In particular, $\langle Y_n \rangle = n\,\mu$ and $\operatorname{Var}[Y_n] = n\,\sigma_{\rm tot}^2$, where $\sigma_{\rm tot}^2=\Psi''(0)$ includes both Gaussian and jump contributions. For coarse-grained dissipation at scale $s$, these cumulants are therefore proportional to the effective depth $n(s)$. They depend linearly on $\log(S/s)$ only in the standard case where $n(s)=\log(S/s)/\log r_\mathrm{c}$; another dependence of $n(s)$ produces a corresponding departure from logarithmic behaviour.

The same observation applies to spatial correlations. Two points separated by a distance $\Delta$ share a common cascade history over a depth $n(\Delta)$ before their branches separate. The additive cascade structure therefore gives
\begin{equation}
\operatorname{Cov}\!\left[\log \varepsilon(0), \log \varepsilon(\Delta)\right] \propto n(\Delta).
\end{equation}
The familiar logarithmic correlation~\cite{arneodo1998analysis} $\operatorname{Cov}[\log\varepsilon(0),\log\varepsilon(\Delta)]\propto\log(S/\Delta)/\log r_\mathrm{c}$ is recovered when the common depth is logarithmic in the separation. Thus, the fundamental signature of log-divisibility is linearity with cascade depth, rather than necessarily with the logarithm of the physical scale.

The cascade also admits a large-deviation description. The depth-normalized exponent
\begin{equation}
\alpha_n = -\frac{1}{n}\log(\varepsilon_n/\varepsilon_0) = -\frac{1}{n} {\textstyle \sum_{i=1}^{n}\mathcal{X}_i}.
\label{eq:def_alpha}
\end{equation}
is an empirical average of independent and identically distributed random variables. It therefore satisfies a large-deviation principle~\cite{touchette2009large}
\begin{equation}
p(\alpha;s) \asymp \exp\!\left[-n(s)\,I(\alpha)\right].
\label{eq:large_dev}
\end{equation}
The rate function can then be written
\begin{equation}
I(\alpha) = \sup_q \left[ q\,\alpha - \Psi(q) \right].
\end{equation}
In the standard scale-invariant case $n(s)=\log(S/s)/\log r_\mathrm{c}$, the exponent $\alpha_n$ coincides with the usual scale-dependent exponent $\alpha_s = [\log(\varepsilon_s/\varepsilon_S)]/[\log(s/S)]$ and the large-deviation form \eqref{eq:large_dev} reduces to the scale-invariant form.
\begin{equation}
p(\alpha;s)
\asymp
(s/S)^{I(\alpha)}.
\label{eq:multifractal_probability}
\end{equation}
In that case, the same function $I(\alpha)$ is directly the multifractal co-dimension spectrum: along a one-dimensional cut through the flow, the set of points characterized by exponent $\alpha$ has fractal dimension $1-I(\alpha)$~\cite{Frisch_1995,meneveau1987multifractal,meneveau1991multifractal}. The corresponding dissipation moments behave as
\begin{equation}
\left\langle\left(\varepsilon_s/\varepsilon_S\right)^q \right\rangle \propto (s/S)^{\chi_q},
\quad\text{with } \chi_q = -\Psi(q)/\log r_\mathrm{c},
\end{equation}
where intermittency manifests itself through the nonlinear, concave dependence of $\chi_q$ on the order $q$.

It is therefore important to distinguish between log-divisibility, large deviations, and multifractality. Log-divisibility is a property of the stochastic evolution in cascade depth and implies that logarithmic cumulants and correlations are proportional to the corresponding depth. A large-deviation principle describes the fluctuations accumulated over large depths, independently of how that depth is related to physical scale. Multifractal power-law scaling is obtained only when the effective depth is asymptotically proportional to $\log(S/s)$. For example, a depth proportional to $[\log(s/s_{\rm c})]^2$ may still yield a well-defined large-deviation function $I(\alpha)$, but not conventional multifractal scaling in $s$. Conversely, multifractal moment scaling does not by itself establish that the underlying increments are independent and stationary, and therefore does not imply log-divisibility.

The following sections examine the effective cascade depth governing Eulerian and Lagrangian coarse-grained dissipation and test whether their statistics are consistent with an underlying scale-invariant multiplicative cascade.

%%%%%%%%%%%%%%%%%%%%%%%%%%%%%%%%%%%%%%
\begin{figure*}[t!]
\includegraphics[width=\linewidth]{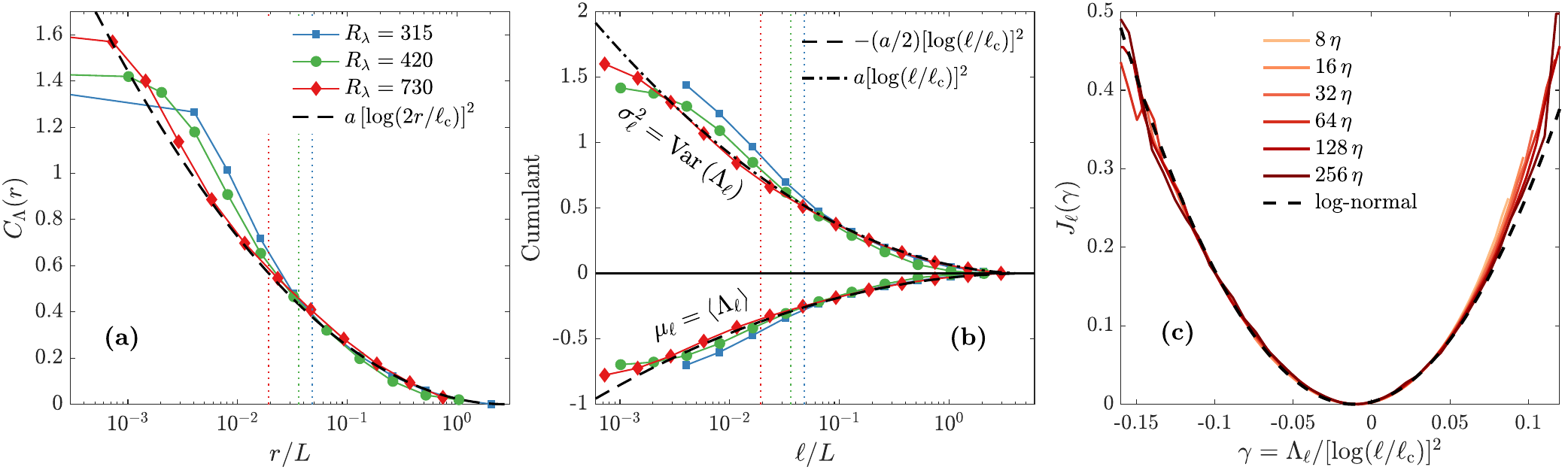}
\caption{\textbf{(a)} Spatial covariance of $\Lambda(\bx,t)=\log[\varepsilon(\bx,t)/\beps]$ as a function of the separation $r$. The dashed line is
$a[\log(2r/\ell_{\rm c})]^2$, with $a=0.023$ and $\ell_{\rm c}=5.5L$. \textbf{(b)} Mean $\mu_\ell$ and variance $\sigma_\ell^2$ of $\Lambda_\ell = \log(\varepsilon_\ell/\beps)$ as functions of the coarse-graining scale $\ell$. are quadratic functions of $\log(\ell/\ell_{\rm c})$ with the
same $a$ and $\ell_{\rm c}$ as in~(a). Vertical dotted lines in panels~(a,b) mark the Taylor microscales. \textbf{(c)} Finite-scale rate functions of $\gamma_\ell = \Lambda_\ell/[\log(\ell/\ell_{\rm c}]^2$ for $R_\lambda=730$. The dashed line is $J(\gamma) = (\gamma + a/2)^2/(2a)$.}
\label{fig:stat_logeps_eul}
\end{figure*}
%%%%%%%%%%%%%%%%%%%%%%%%%%%%%%%%%%%%%%

%%%%%%%%%%%%%%%%%%%%%%%%%%%%%%%%%%%%%%
\section{Eulerian dissipation: beyond the scale-invariant spatial cascade}
\label{sec:eulerian}

As discussed in the previous section, two properties commonly associated with multiplicative cascades should be distinguished. A scale-invariant log-divisible cascade requires the logarithm of the coarse-grained dissipation to have stationary increments in logarithmic scale. This entails logarithmic spatial correlations, cumulants growing linearly with the cascade depth, and scale-independent multiplier statistics. Multifractality is a separate property: it requires a large-deviation principle with speed proportional to $\log(L/\ell)$, or equivalently power-law scaling of the moments of the coarse-grained dissipation. We examine these two structures successively and show that neither is observed. Instead, the data reveal a different large-deviation structure governed by a squared-logarithmic scale variable.

We first test the log-divisible cascade picture. In a spatial multiplicative cascade, two points separated by a distance $r=|\boldsymbol{r}|$ share the same cascade history down to a scale of order $r$. The covariance of the logarithmic dissipation should therefore be proportional to the number of common cascade steps and hence to $\log(L/r)$. Introducing $\Lambda(\bx,t)=\log[\varepsilon(\bx,t)/\beps]$, we measure
\begin{equation}
C_\Lambda(r) = \left\langle \Lambda(\bx,t)\Lambda(\bx+\boldsymbol{r},t) \right\rangle - \left\langle\Lambda\right\rangle^2, \qquad r = |\boldsymbol{r}|
\end{equation}
As shown in Fig.~\ref{fig:stat_logeps_eul}(a), $C_\Lambda(r)$ does not display an extended range of linear dependence on $\log r$. At most, such a behavior could be inferred over a short interval below the Taylor microscale, which  broadens only weakly as the Reynolds number increases. Over the resolved range, the covariance is instead much better described by
\begin{equation}
C_\Lambda(r) \approx a\,\left[\log\!\left(2\,r/\ell_{\rm c}\right)\right]^2,
\label{eq:cov_logeps_eul}
\end{equation}
with $a\simeq0.023$ and $\ell_{\rm c}\simeq5.5L$. This behavior is incompatible with the logarithmic correlations expected from a scale-invariant spatial cascade~\cite{o1993spatial}.

These observations should be compared with the recent analysis of Ruffenach and Chevillard~\cite{ruffenach2026spatiotemporal}, who reported an approximately logarithmic spatial covariance of $\log\varepsilon$, with $\mu\simeq0.21$, for simulations with much higher Reynolds numbers compared to ours.
Their logarithmic form $\mu\log(L^\star/r)$ introduces an effective length $L^\star$ that absorbs an additive finite-scale correction, while the slope is fixed to the reported value $\mu\simeq0.21$. Over a finite range, the two forms are therefore more sharply distinguished by their local slope than by the quality of the fit to $C_\Lambda$ alone. Equation~\eqref{eq:cov_logeps_eul} implies a scale-dependent effective coefficient
\begin{equation}
\mu_{\rm eff}(r)=-\frac{\mathrm d C_\Lambda}{\mathrm d\log r}
=2a\log\!\left(\frac{\ell_{\rm c}}{2r}\right),
\end{equation}
whereas a logarithmic covariance requires a constant slope. With the values of $a$ and $\ell_{\rm c}$ given above, $\mu_{\rm eff}$ decreases systematically across the resolved range, from approximately $0.22$ near the Taylor microscale to about $0.05$ near $L$ for $R_\lambda=730$. Thus, over a restricted interval concentrated around and below $\lambda$, a logarithmic fit can return a value close to the conventional $\mu\simeq0.21$, even though the local slope continues to vary across the broader range. The present analysis differs in that the same pair $(a,\ell_{\rm c})$ also constrains the one-point cumulants, the rate function, and the moments of $\varepsilon_\ell$, providing a joint constraint that a covariance fit alone does not test. Whether the broader logarithmic range reported in 
Ref.~\cite{ruffenach2026spatiotemporal} at higher Reynolds numbers represents an asymptotic logarithmic regime or instead a slowly varying effective slope can therefore be tested directly through the local-slope behavior of $C_\Lambda$.

%%%%%%%%%%%%%%%%%%%%%%%%%%%%%%%%%%%%%%
\begin{figure*}[t!]
	\includegraphics[width=\linewidth]{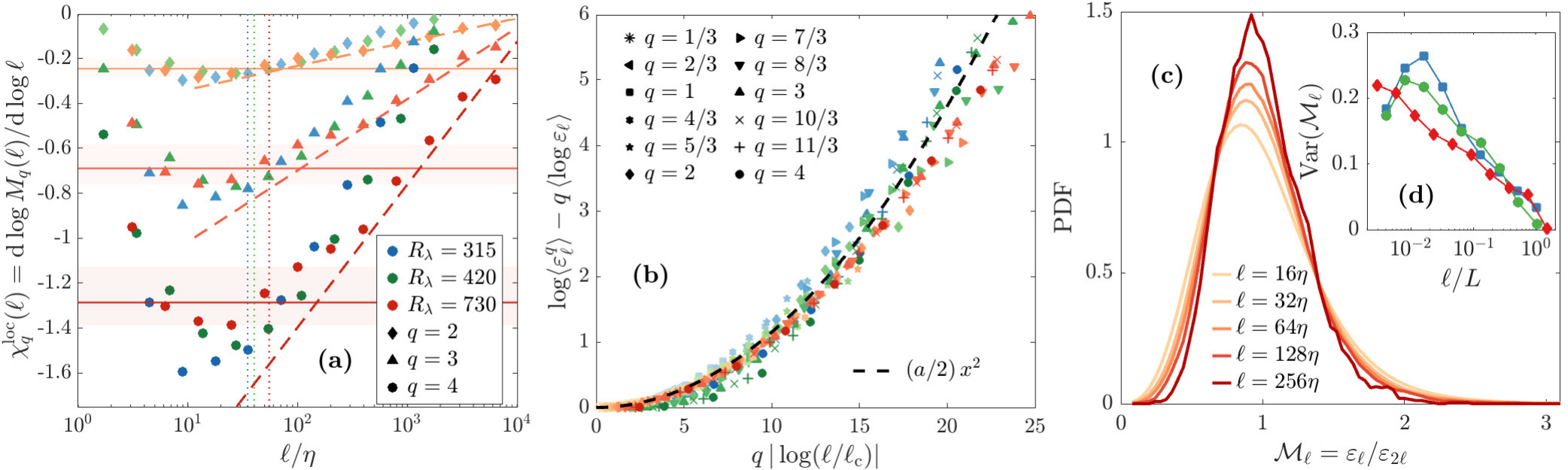}
\caption{\textbf{(a)} Local slopes $\chi_q^{\rm loc}(\ell)=\dd\log M_q/\dd\log\ell$ of $M_q(\ell)=\langle(\varepsilon_\ell/\beps)^q\rangle$. Horizontal lines and shaded bands show estimates of the scaling exponents $\chi_q$ over $\eta\lesssim\ell\lesssim\lambda$ for the three Reynolds numbers; dashed lines follow from \eqref{eq:local_slopes_eul}. \textbf{(b)} Centered cumulant-generating function plotted against $q|\log(\ell/\ell_{\rm c})|$ for $\ell\in[2\eta,2000\eta]$ and $q\leq4$. The dotted line is the quadratic form $a\,x^2/2$. \textbf{(c)} PDFs of the Eulerian multipliers $\mathcal{M}_\ell=\varepsilon_{\ell}/\varepsilon_{2\ell}$ for $R_\lambda = 730$. \textbf{(d)} Variance of $\mathcal{M}_\ell$ for the three Reynolds numbers.}
\label{fig:mom_eps_eul}
\end{figure*}
%%%%%%%%%%%%%%%%%%%%%%%%%%%%%%%%%%%%%%

A complementary test is provided by the cumulants of the coarse-grained logarithmic dissipation
\begin{equation}
\Lambda_\ell = \log(\varepsilon_\ell/\beps), \qquad \mu_\ell=\langle\Lambda_\ell\rangle, \qquad \sigma_\ell^2={\rm Var}(\Lambda_\ell).
\end{equation}
Let $n=\log_2(L/\ell)$ denote the number of dyadic scale steps between $L$ and $\ell$. For a log-divisible cascade with stationary increments in $n$, every cumulant of $\Lambda_\ell$ must be proportional to $n$. In particular, both $\mu_\ell$ and $\sigma_\ell^2$ should depend linearly on $\log(L/\ell)$. Figure~\ref{fig:stat_logeps_eul}(b) shows a markedly different behavior. For all three Reynolds numbers, the mean and variance are well approximated by quadratic functions,
\begin{equation}
\mu_\ell\approx -(a/2)\,[\log(\ell/\ell_{\rm c})]^2, \qquad \sigma_\ell^2\approx a [\log(\ell/\ell_{\rm c})]^2,
\label{eq:quadratic_mu_sigma}
\end{equation}
with the same values of $a$ and $\ell_{\rm c}$ as in \eqref{eq:cov_logeps_eul}. The logarithm of the Eulerian coarse-grained dissipation therefore cannot be represented as a process with stationary additive increments in the usual logarithmic scale variable.

The agreement between Figs.~\ref{fig:stat_logeps_eul}(a) and~(b) is, however, noteworthy. Upon identifying the coarse-graining scale with $\ell\simeq2r$, the one-point cumulants and the two-point covariance are governed by the same function
\begin{equation}
n(\ell) := \left[ \log(\ell/\ell_{\rm c}) \right]^2.
\label{eq:nonlinear_cascade_depth}
\end{equation}
This suggests that $n(\ell)$ acts as an effective, nonlinear cascade depth. The data thus retain a coherent organization across scales, but its clock is quadratic rather than linear in the logarithmic scale separation. This observation motivates introducing the scale-averaged logarithmic
dissipation
\begin{equation}
\gamma_\ell := \frac{\Lambda_\ell}{n(\ell)} = \frac{\log(\varepsilon_\ell/\beps)}{[\log(\ell/\ell_{\rm c})]^2}.
\label{eq:def_loginc}
\end{equation}
Figure~\ref{fig:stat_logeps_eul}(c) shows finite-scale estimates of the associated rate function,
\begin{equation}
J_\ell(\gamma) = -\frac{1}{n(\ell)} \log\!\left[ \frac{p_{\gamma_\ell}(\gamma;\ell)}{p_{\gamma_\ell}(\gamma_\ell^\star;\ell)}\right],
\label{eq:finite_rate_eul}
\end{equation}
where $\gamma_\ell^\star$ denotes the most probable value, so that $\min J_\ell=0$. The curves obtained at different scales collapse closely over more than one decade in $\ell$. This supports the large-deviation form
\begin{equation}
p_{\gamma_\ell}(\gamma;\ell) \asymp \exp\!\left[-n(\ell)\,J(\gamma)\right], \qquad \eta\ll\ell\ll L,
\label{eq:scaling_pdf_epsil_eul}
\end{equation}
where equality is understood at logarithmic accuracy.

The measured rate function is remarkably close to the parabola
\begin{equation}
J(\gamma) \approx \frac{(\gamma+a/2)^2}{2a}.
\label{eq:parabolic_rate_eul}
\end{equation}
The Eulerian dissipation therefore displays an approximately log-normal large-deviation structure, but with the nonstandard speed $n(\ell)=[\log(\ell/\ell_{\rm c})]^2$. This differs fundamentally from both a log-normal multiplicative cascade and the conventional multifractal formalism, for which the large-deviation speed is proportional to $\log(L/\ell)$.

We next examine the implications of this nonstandard large-deviation principle for the moments of the coarse-grained dissipation. Define the normalized moments $M_q(\ell) =\langle(\varepsilon_\ell/\beps)^q\rangle$ and their logarithmic derivatives $\chi_q^{\rm loc}(\ell) = \dd\log M_q(\ell)/\dd\log\ell$. If the dissipation were multifractal over an inertial range, $M_q(\ell)\propto (\ell/L)^{\chi_q}$, then $\chi_q^{\rm loc}(\ell)$ would develop a scale-independent plateau. As seen in Fig.~\ref{fig:mom_eps_eul}(a), no extended plateau is observed for $q=2$, $3$, or $4$, except possibly over the restricted interval $\eta \lesssim \ell\lesssim\lambda$ below the Taylor micro-scale, where the local slopes can be approximated by effective constants. At $R_\lambda=730$, this gives $\chi_2\simeq-0.25$, $\chi_3\simeq-0.69$, and $\chi_4\simeq-1.29$, values that can be made compatible with familiar log-normal or log-Poisson estimates. These effective exponents are not, however, maintained over a substantial range of scales. Above the Taylor microscale, the local slopes drift approximately linearly with $\log\ell$, and comparison among the three Reynolds numbers reveals no systematic tendency for scale-independent plateaux to emerge.

This drift follows directly from the large-deviation structure identified above. By the Laplace principle,
\begin{equation}
\frac{1}{n(\ell)}\log M_q(\ell) \longrightarrow \psi(q) := \sup_\gamma\{q\gamma-J(\gamma)\},
\label{eq:scgf_eul}
\end{equation}
where $\psi$ is the scaled cumulant-generating function. For the parabolic rate function \eqref{eq:parabolic_rate_eul},
\begin{equation}
\psi(q) = (a/2)\,(q^2-q).
\end{equation}
The linear term ensures $\psi(1)=0$, consistently with $M_1(\ell)=1$. The moments therefore obey
\begin{equation}
\log M_q(\ell) \approx (a/2)\,(q^2-q) \left[ \log(\ell/\ell_{\rm c})\right]^2.
\label{eq:scaling_mom_epsil_eul}
\end{equation}
This is a squared-logarithmic scaling law, rather than a power law in $\ell$.

Figure~\ref{fig:mom_eps_eul}(b) tests this prediction directly. It displays the centered cumulant-generating function
\begin{equation}
K(q;\ell) := \log M_q(\ell)-q\mu_\ell = \log \left\langle \exp\!\left[q(\Lambda_\ell-\mu_\ell)\right] \right\rangle.
\end{equation}
Equations~\eqref{eq:quadratic_mu_sigma} and \eqref{eq:parabolic_rate_eul} predict
\begin{equation}
K(q;\ell) \approx (a/2)\,q^2 n(\ell) =(a/2) \left[ q\log(\ell/\ell_{\rm c})\right]^2.
\label{eq:centered_cgf_eul}
\end{equation}
The collapse of the data onto the parabola $(a/2)\,x^2$ confirms this prediction over the resolved scales and for moments up to $q=4$.

Differentiating \eqref{eq:scaling_mom_epsil_eul} gives
\begin{equation}
\chi_q^{\rm loc}(\ell) \approx a(q^2-q) \log(\ell/\ell_{\rm c}).
\label{eq:local_slopes_eul}
\end{equation}
The dashed lines in Fig.~\ref{fig:mom_eps_eul}(a) show this prediction. They account for the approximately linear evolution of the local slopes throughout the upper part of the inertial range. Their drift is therefore not merely due to the absence of a sufficiently broad fitting interval: it is the moment-space signature of the squared-logarithmic large-deviation speed.

The failure of a self-similar Eulerian cascade is also apparent directly at the level of the multipliers. For each reference scale $\ell$, define
\begin{equation}
\mathcal{M}_\ell = \varepsilon_\ell/\varepsilon_{2\ell}, \qquad \log\mathcal{M}_\ell = \Lambda_\ell-\Lambda_{2\ell}.
\end{equation}
Stationarity of the increments in logarithmic scale would require the distribution of $\mathcal{M}_\ell$ to be independent of $\ell$. Figure~\ref{fig:mom_eps_eul}(c) instead shows a systematic evolution of the multiplier PDFs with the reference scale. Their widths first increase toward the dissipative range and then decrease continuously as $\ell$ approaches the integral scale. The variance shown in Fig.~\ref{fig:mom_eps_eul}(d) confirms this strong scale dependence for all three Reynolds numbers. The curves become close in the upper part of the inertial range, but retain a Reynolds-number dependence at the smallest resolved scales, where viscous effects are expected. Most importantly, increasing $R_\lambda$ does not produce a scale-independent plateau. The multiplier increments therefore remain nonstationary across scales. The observed scale dependence does not exclude more general multiplicative descriptions involving nonstationary or scale-dependent multiplier statistics \citep{castaing1995fully,arneodo1998analysis}. Such constructions, however, depart from the scale-invariant cascade phenomenology summarized in Sec.~\ref{sec:multifractal}. More specifically, the present results point to an effective cascade depth $n(\ell)=[\log(\ell/\ell_{\rm c})]^2$, whose physical origin is not accounted for by conventional cascade models.

These observations do not constitute a statement about the formal limit $R_\lambda\to\infty$. They do, however, show that no conventional multifractal scaling regime is resolved in the present simulations. At the largest Reynolds number, the separation between the integral and Taylor scales reaches approximately $L/\lambda\simeq50$, yet no extended plateau develops. Moreover, neither the local slopes nor the scale dependence of the logarithmic cumulants or multiplier distributions shows a systematic evolution toward the expectations of a scale-invariant cascade. 

Taken together, Figs.~\ref{fig:stat_logeps_eul} and \ref{fig:mom_eps_eul} establish a coherent but nonclassical organization of Eulerian dissipation. Its logarithm does not have stationary increments in $\log\ell$, and its moments do not obey multifractal power laws. Instead, the one-point cumulants, spatial covariance, rate function, and moments are all governed by the effective depth $n(\ell)=[\log(\ell/\ell_{\rm c})]^2$, with an approximately parabolic rate function. Eulerian intermittency thus retains a large-deviation structure, but one whose speed is incompatible with the scale-invariant spatial cascade phenomenology usually invoked in turbulence.

The emergence of this squared-logarithmic depth is not a minor correction to the conventional picture. It implies that the effective amount of statistical accumulation between two scales is not proportional to their logarithmic separation, as it would be for a self-similar hierarchy of scale-local cascade steps. Its physical origin therefore remains to be identified. It reflects non-stationary transfer statistics, correlations extending across scales, or the projection onto space of dynamical events whose natural organization is temporal rather than spatial. Since dissipation is experienced along fluid trajectories, this observation motivates examining whether a more conventional cascade structure emerges in a Lagrangian description.
%%%%%%%%%%%%%%%%%%%%%%%%%%%%%%%%%%%%%%
\begin{figure*}[t!]
\includegraphics[width=\linewidth]{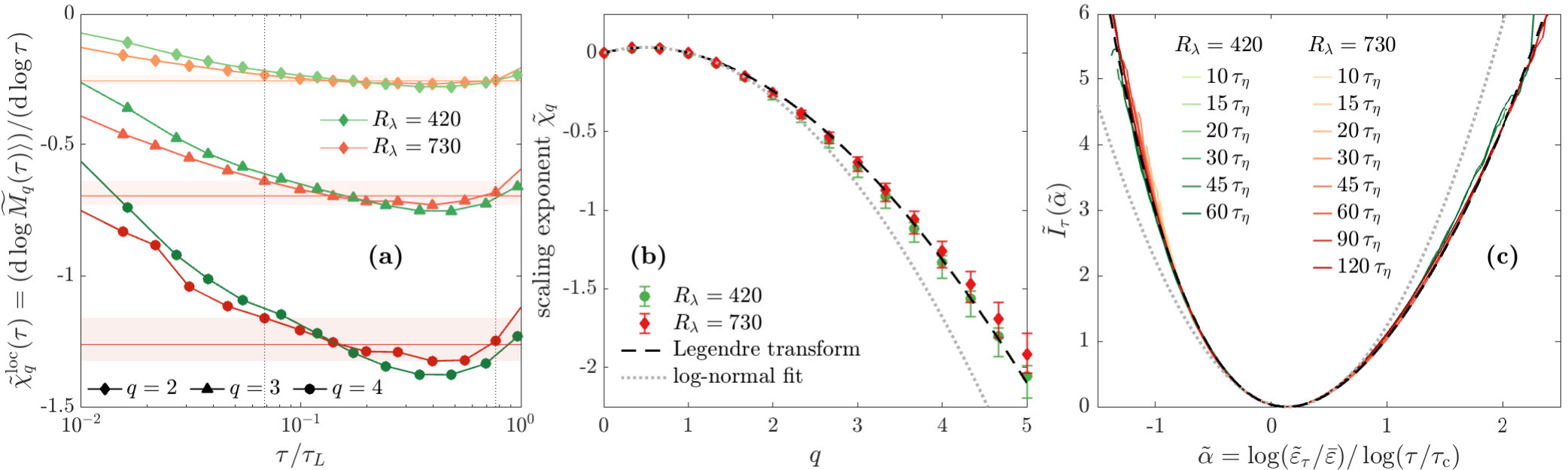}
\caption{\textbf{(a)} Local slopes $\tilde{\chi}^{\rm loc}_q(\tau)=\dd\log\widetilde{M}_q(\tau)/\dd\log\tau$ of the moments $\widetilde{M}_q(\tau) =\langle(\tilde{\varepsilon}_\tau/\beps)^q\rangle$. The vertical dotted lines delimit the interval $8\tau_\eta<\tau<0.4\tau_L$ used to estimate the scaling exponents, shown as horizontal lines. The shaded regions indicate the corresponding uncertainties, determined from the extrema of the local slopes over this interval. \textbf{(b)} Measured scaling exponents $\tilde{\chi}_q$, with associated error bars. The dashed curve is obtained from the Legendre transform of the quartic fit to the rate function shown in~(c), while the dotted curve is the quadratic log-normal prediction $\tilde{\chi}_q=0.14\,q(1-q)$. \textbf{(c)} Finite-scale estimates of the temporal rate function $\tilde{I}(\tilde{\alpha})$, with $\tilde{\alpha} =\log(\tilde{\varepsilon}_\tau/\beps)/\log(\tau/\tau_{\rm c})$, for the indicated coarse-graining times and Reynolds numbers. The dashed curve is a quartic fit to the collapsed data, while the dotted curve is the quadratic rate function associated with the log-normal prediction in panel~(b).}
\label{fig:mom_eps_lag}
\end{figure*}
%%%%%%%%%%%%%%%%%%%%%%%%%%%%%%%%%%%%%%

%%%%%%%%%%%%%%%%%%%%%%%%%%%%%%%%%%%%%%
\section{Multifractality and temporal cascade of Lagrangian dissipation}
\label{sec:lagrangian}

\begin{figure*}[t]
\includegraphics[width=\linewidth]{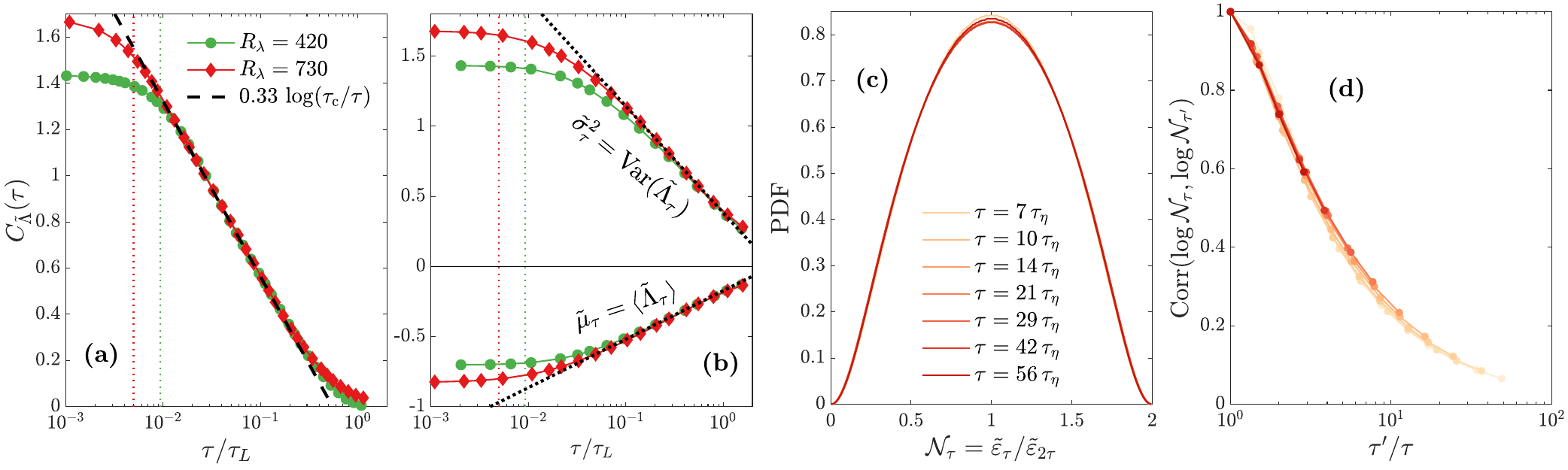}
\caption{\textbf{(a)} Temporal covariance of $\tilde{\Lambda}(t)=\log[\tilde{\varepsilon}(t)/\beps]$. The dashed line shows the logarithmic behavior $0.33\log(\tau_{\ast}/\tau)$. \textbf{(b)}~Mean $\tilde{\mu}_\tau$ and variance $\tilde{\sigma}_\tau^2$ of $\tilde{\Lambda}_\tau =\log(\tilde{\varepsilon}_\tau/\beps)$. Dotted lines indicate linear dependences on $\log(\tau/\tau_{\rm c})$. \textbf{(c)}~PDFs of the temporal multipliers $\mathcal{N}_\tau=\tilde{\varepsilon}_{\tau}/\tilde{\varepsilon}_{2\tau}$, for $R_\lambda=730$ and the indicated coarse-graining times. \textbf{(d)} Correlation coefficients $\operatorname{Corr}(\log\mathcal{N}_\tau,\log\mathcal{N}_{\tau'})$ between logarithmic multipliers, as functions of the timescale ratio $\tau^\prime/\tau$, for the same reference times $\tau$ as in panel~(c).}
\label{fig:stat_logeps_lag}
\end{figure*}

We now turn to the dissipation sampled along fluid trajectories. Tracer particles evolve according to $\dd\bX/\dd t = \bv(\bX,t)$, and the instantaneous Lagrangian dissipation is defined as $\tilde{\varepsilon}(t) = \varepsilon(\bX(t),t)$. For tracers uniformly distributed in an incompressible flow, the one-time probability distribution of $\tilde{\varepsilon}$ coincides with the single-point Eulerian distribution of $\varepsilon$. Any difference between the two descriptions must therefore originate from the correlations and scale organization of the fluctuations, rather than from their instantaneous statistics. To probe this temporal organization, we introduce the dissipation averaged along a trajectory over a time interval $\tau$,
\begin{equation}
    \tilde{\varepsilon}_\tau(t) = \frac{1}{\tau} \int_t^{t+\tau} \tilde{\varepsilon}(t')\,\dd t'.
\label{eq:def_eps_lag}
\end{equation}
We use the large-time cutoff $\tau_{\rm c}=3.2\,\tau_L$, as determined below from the logarithmic scaling of the mean and variance of
$\log(\tilde{\varepsilon}_\tau/\beps)$ [Fig.~\ref{fig:stat_logeps_lag}(b)].

We first examine whether $\tilde{\varepsilon}_\tau$ displays multifractal scaling. Its normalized moments are $\widetilde{M}_q(\tau) = \langle ( \tilde{\varepsilon}_\tau/\beps )^q \rangle$. Figure~\ref{fig:mom_eps_lag}(a) shows their local logarithmic slopes $\tilde{\chi}^{\rm loc}_q(\tau) = (\dd\log\widetilde{M}_q(\tau))/(\dd\log\tau)$. For $q=2,3,4$, the slopes develop broad plateaux over the interval $8\tau_\eta<\tau<0.4\tau_L$. These plateaux are observed at both Reynolds numbers and become more clearly defined as $R_\lambda$ increases. This behavior sharply contrasts with the Eulerian results of Sec.~\ref{sec:eulerian}, where the local slopes retain a pronounced scale dependence, and provides direct evidence for the inertial-range power laws $\widetilde{M}_q(\tau) \propto (\tau/\tau_{\rm c})^{\tilde{\chi}_q}$.

The resulting exponents are reported in Fig.~\ref{fig:mom_eps_lag}(b). The measurements at the two Reynolds numbers agree within their uncertainties and define a nonlinear concave function of $q$. The constraints $\tilde{\chi}_0=0$ and $\tilde{\chi}_1=0$ follow, respectively, from normalization and from $\langle\tilde{\varepsilon}_\tau\rangle=\beps$. The nonlinear dependence on $q$ establishes the multifractal character of the temporally coarse-grained dissipation. A quadratic approximation, $\tilde{\chi}^{\rm LN}_q=0.14\,q(1-q)$, provides a reasonable description at low orders, but predicts increasingly negative exponents as $q$ increases and thus systematically overestimates their magnitude. The Lagrangian dissipation is therefore multifractal, but its statistics are not accurately described by a log-normal model.

The same conclusions follows from the complementary large-deviation formulation. Multifractal scaling implies
\begin{equation}
    p(\tilde{\alpha};\tau) \asymp (\tau/\tau_{\rm c})^{\tilde{I}(\tilde{\alpha})},
    \qquad \tilde{\alpha} = \frac{\log(\tilde{\varepsilon}_\tau/\beps)}{\log(\tau/\tau_{\rm c})},
\label{eq:large_dev_eps_lag}
\end{equation}
where $\tilde{I}(\tilde{\alpha})$ is a convex temporal rate function normalized by $\inf_{\tilde{\alpha}}\tilde{I}(\tilde{\alpha})=0$. The moment exponents are related to it through $\tilde{\chi}_q = \inf_{\tilde{\alpha}} [q\,\tilde{\alpha}+\tilde{I}(\tilde{\alpha})]$. Figure~\ref{fig:mom_eps_lag}(c) shows the finite-scale estimates of $\tilde{I}(\tilde{\alpha})$. Results obtained at different coarse-graining times collapse closely onto a common curve, with only weak dependence on Reynolds number. The measured rate function is clearly non-quadratic.  Its Legendre transform, obtained from the quartic fit shown by the dashed curve in Fig.~\ref{fig:mom_eps_lag}(c), is reported as the dashed curve in panel~(b) and reproduces the measured exponents, including their departure from the quadratic log-normal prediction. The local-slope plateaux, the nonlinear exponents, and the collapse of the rate functions thus provide mutually consistent evidence for non-log-normal multifractality.

%%%%%%%%%%%%%%%%%%%%%%%%%%%%%%%%%%%%%%
%%%%%%%%%%%%%%%%%%%%%%%%%%%%%%%%%%%%%%

Similar inertial-range scaling and logarithmic statistics of the Lagrangian dissipation were previously reported at $R_\lambda=400$ by Huang \& Schmitt~\cite{huang2014lagrangian}. Their results were found to be approximately compatible with a log-normal model. The present data, obtained at higher Reynolds numbers and with improved statistical resolution, confirm the multifractal scaling of the Lagrangian coarse-grained dissipation while revealing systematic departures from log-normality.

We next ask whether this multifractal scaling is accompanied by the statistical organization expected from a multiplicative process. We first consider the logarithm of the instantaneous dissipation, $\tilde{\Lambda}(t) = \log[\tilde{\varepsilon}(t)/\beps]$. Its connected temporal covariance,
\begin{equation}
    C_{\tilde{\Lambda}}(\tau) = \langle \tilde{\Lambda}(t)\,\tilde{\Lambda}(t+\tau) \rangle - \langle\tilde{\Lambda}\rangle^2,
\label{eq:cov_logeps_lag}
\end{equation}
is shown in Fig.~\ref{fig:stat_logeps_lag}(a). Over most of the inertial range, it follows
\begin{equation}
    C_{\tilde{\Lambda}}(\tau) \simeq c_2\log(\tau_{\ast}/\tau), \quad c_2\simeq0.33,\ \tau_{\ast}\simeq0.55\,\tau_L.
\label{eq:log_cov_lag}
\end{equation}
This logarithmic decay is the temporal counterpart of the logarithmic spatial correlations predicted by scale-invariant multiplicative cascades. Its clarity is particularly striking in view of the absence of an analogous regime in the Eulerian dissipation field.

Further evidence is provided by the logarithm of the temporally averaged dissipation, $\tilde{\Lambda}_\tau = \log(\tilde{\varepsilon}_\tau/\beps)$.  As shown in Fig.~\ref{fig:stat_logeps_lag}(b), both its mean $\tilde{\mu}_\tau=\langle\tilde{\Lambda}_\tau\rangle$ and its variance $\tilde{\sigma}_\tau^2=\operatorname{Var}(\tilde{\Lambda}_\tau)$ vary linearly with the logarithm of the coarse-graining time,
\begin{equation}
    \tilde{\mu}_\tau \simeq c_1\log(\tau/\tau_{\rm c}), \qquad \tilde{\sigma}_\tau^2 \simeq c_2\log(\tau_{\rm c}/\tau).
\label{eq:log_cumulants_lag}
\end{equation}
The two linear dependences consistently extrapolate to zero at $\tau_{\rm c}\simeq3.2\,\tau_L$, thereby fixing the large-time cutoff used above in the definition of $\tilde{\alpha}$. In particular, $\tau_{\rm c}$ is determined independently of the collapse of the rate functions in Fig.~\ref{fig:mom_eps_lag}(c). The variance has the same logarithmic slope $c_2\simeq 0.33$ as the covariance of the instantaneous signal. For a scale-invariant log-divisible process, the cumulants of $\tilde{\Lambda}_\tau$ are expected to depend linearly on $\log(\tau/\tau_{\rm c})$. Equation~\eqref{eq:log_cumulants_lag} tests this prediction for the first two cumulants. This behavior again differs markedly from the quadratic dependence found under Eulerian spatial coarse-graining.

A direct scale-by-scale test is provided from the temporal multipliers
\begin{equation}
    \mathcal{N}_\tau = \tilde{\varepsilon}_{\tau}/\tilde{\varepsilon}_{2\tau}.
\label{eq:def_lag_multipliers}
\end{equation}
Their logarithms are the increments of $\tilde{\Lambda}_{\tau}$ between timescales separated by a fixed ratio of two. As shown in Fig.~\ref{fig:stat_logeps_lag}(c), the probability distributions of $\mathcal{N}_\tau$ collapse closely for the various values of $\tau$ within the inertial range. The multiplier statistics are therefore approximately stationary with respect to translation of the logarithmic timescale.

Scale invariance of the multiplier distributions does not imply that successive scale increments are independent. We therefore consider $\rho(\tau,\tau^\prime) = \operatorname{Corr}(\log\mathcal{N}_\tau,\log\mathcal{N}_{\tau^\prime})$. As shown in Fig.~\ref{fig:stat_logeps_lag}(d), $\rho(\tau,\tau^\prime)$ depends only weakly on the reference scale $\tau$ and decays primarily with the ratio $\tau^\prime/\tau$. The correlations remain measurable over scale ratios extending to several tens. Because the multipliers are constructed from nested temporal averages, part of this correlation may reflect the overlap of the averaging intervals. The measurements of Fig.~\ref{fig:stat_logeps_lag}(d) nevertheless show that the logarithmic scale increments cannot be treated as independent. They are approximately stationary in logarithmic timescale, but retain correlations over several scales.

Taken together, the logarithmic covariance, the linear dependence of the first two logarithmic cumulants, and the scale invariance of the multiplier distributions provide strong evidence for an approximately scale-stationary multiplicative process in time. These observations support a log-divisible description of the single-scale distributions. The correlations in Fig.~\ref{fig:stat_logeps_lag}(d), however, rule out the simplest cascade of independent multipliers and instead reveal a correlated multiplicative organization with memory extending over several temporal scales.

The particularly clear multiplicative organization observed in the Lagrangian frame may be related to the dynamics of material deformation~\cite{meneveau2011lagrangian,yu2010lagrangian,JohnsonWilczekReview2024}. Let
$J_{ij}(t,t_0)=\partial X_i(t)/\partial X_j(t_0)$ denote the deformation gradient along a tracer trajectory. It satisfies $\partial_t{J}_{ij} = A_{ik}\,J_{kj}$ with $A_{ij}=\partial_jv_i$. This leads to rewriting the velocity gradient as $\mathbb{A} = \partial_t{\mathbb{J}}\,\mathbb{J}^{-1}$, so that the dissipation can be express exactly as 
\begin{equation}
    \tilde{\varepsilon}(t) = \frac{\nu}{2} \left| (\partial_t{\mathbb{J}}\,\mathbb{J}^{-1}) + (\partial_t{\mathbb{J}}\,\mathbb{J}^{-1})^\top\right|^2
\label{eq:diss_deformation_lag}
\end{equation}
It therefore measures the squared instantaneous rate of material deformation.

Finite-time deformations compose multiplicatively, $\mathbb{J}(t_2,t_0) = \mathbb{J}(t_2,t_1)\,\mathbb{J}(t_1,t_0)$, so that a fluid trajectory retains the causal succession of stretching and compression events experienced by the same material element. Equation~\eqref{eq:diss_deformation_lag} does not imply that the dissipation is itself a multiplicative functional of $\mathbb{J}$; rather, it shows that its instantaneous fluctuations are generated within a material history whose finite-time deformations compose multiplicatively.

The inter-timescale correlations observed in Fig.~\ref{fig:stat_logeps_lag}(d) may reflect the fluctuating lifetime of Lagrangian gradient-amplification events. Indeed, the quadratic self-amplification term in the velocity-gradient dynamics introduces a local evolution time of order
$\tau_{\rm loc}\sim|\mathbb{A}|^{-1}$. At dissipative scales, this reduces, up to factors of order unity, to the local Kolmogorov time $\tau_{\rm loc} \sim \sqrt{\nu/\tilde{\varepsilon}}$, while for coarse-grained gradients it is of the order of the corresponding local eddy-turnover time. A coherent amplification event may therefore contribute to several neighbouring temporal averages and correlate their associated multipliers. The material history sampled along a Lagrangian trajectory can thus retain memory across several temporal scales, rather than reducing to a sequence of independent cascade steps.

This material genealogy is lost under Eulerian spatial coarse-graining. Nested Eulerian regions contain fluid elements with different deformation histories and therefore mix distinct sequences of strain amplification. The preservation of a single causal history along each trajectory provides a plausible explanation for why the Lagrangian statistics recover the standard organization of a scale-invariant temporal cascade, with a cascade depth proportional to $\log(\tau/\tau_{\rm c})$. This interpretation is consistent with causal multifractal processes and with Lagrangian velocity-gradient models based on recent material deformation~\cite{schmitt2003causal,chevillard2006lagrangian,meneveau2011lagrangian,pereira2018multifractal}. It should not be viewed as a derivation of the observed temporal cascade from the Navier--Stokes equations, but rather as a dynamical rationale for its emergence.

The contrast with the Eulerian results of Sec.~\ref{sec:eulerian} is therefore not simply between multifractal and non-multifractal statistics. Both descriptions exhibit a large-deviation structure, but with fundamentally different scale depths. In the Lagrangian frame, the rate function is organized by the usual logarithmic depth $n(\tau)\propto\log(\tau/\tau_{\rm c})$, consistently with logarithmic correlations and approximately scale-stationary multipliers. In the Eulerian frame, by contrast, the large-deviation collapse requires the non-standard depth $n(\ell)\propto[\log(\ell/\ell_{\rm c})]^2$, while logarithmic spatial correlations and scale-independent multipliers are absent. The bridge between the two descriptions must therefore account not only for the conversion of spatial scales into temporal ones, but also for this change in the effective cascade depth. Classical bridge relations~\cite{borgas1993multifractal,chevillard2012phenomenological} are re-examined from this perspective in the following section.

%%%%%%%%%%%%%%%%%%%%%%%%%%%%%%%%%%%%%%

\section{Revisiting bridge relations: from spatial sampling to temporal inheritance}
\label{sec:bridge}

Bridge relations seek to connect the spatial organization of turbulence to the temporal fluctuations experienced along particle trajectories. For velocity increments, dimensional conversions between spatial separations and eddy-turnover times have proved remarkably successful, yielding Eulerian and Lagrangian scaling exponents that are broadly consistent with one another~\cite{arneodo2008universal}. Their application to dissipation is more delicate. Indeed, $\varepsilon_\ell$ is an instantaneous spatial average, whereas $\tilde{\varepsilon}_\tau$ records the history of dissipation along a moving particle. Relating these two quantities therefore requires assumptions not only about the conversion between $\ell$ and $\tau$, but also about how dissipation is sampled, transported, and inherited by material trajectories.
	
\begin{figure}[h]
    \resizebox{0.9\linewidth}{!}{\begin{tikzpicture}[x=1cm,y=1cm,line cap=round,line join=round,
timearrow/.style={
    black!65,
    line width=0.95pt,
    -{Stealth[length=2.8mm,width=2.0mm]}
  },
material/.style={
    draw=flowblue,
    dash pattern=on 4pt off 2.5pt,
    line width=1.45pt
  }]

% ------------------------------------------------------------------
% Sweeping panel
% ------------------------------------------------------------------
\path[fill=blobdark]
  plot[smooth cycle,tension=0.8] coordinates {
    (0.35,1.55) (0.55,2.55) (1.40,2.95) (2.45,2.75)
    (3.00,2.10) (2.80,1.15) (1.95,0.80) (0.85,0.95)
  };

% Streamlines through the small eddy
\draw[flow]
  plot[smooth,tension=0.65] coordinates {(0.60,2.45) (1.3,2.15) (2.05,2.15) (3.0,2.24)};
\draw[flow]
  plot[smooth,tension=0.65] coordinates {(0.45,2.05) (1.70,1.93) (2.4,1.96) (3.05,2.0)};
\draw[flow]
  plot[smooth,tension=0.65] coordinates {(0.43,1.78) (1.72,1.70) (2.40,1.55) (3.0,1.36)};
\draw[flow]
  plot[smooth,tension=0.65] coordinates {(0.47,1.40) (1.35,1.48) (1.95,1.4) (2.75,1.05)};

\node[epslab] at (1.65,2.55) {$\varepsilon_\ell(t)$};

\draw[dim] (0.50,0.52) -- (2.70,0.52);
\node[lab,below=2pt] at (1.60,0.56) {$\ell$};
\node[font=\Large,above] at (1.55,-1.2) {$\tilde{\varepsilon}_{\tau_\ell} \sim \varepsilon_\ell$};
\node[font=\Large,below] at (1.55,-1.2) {{\bfseries (a)} {\sf Spatial sampling}};

\draw[timearrow] (0.55,3.5) -- (3.15,3.5);
\node[font=\Large,above=2pt] at (0.55,3.5) {$t-\tau_\ell$};
\node[font=\Large,above=2pt] at (3.15,3.5) {$t$};

% ------------------------------------------------------------------
% inheritance panel
% ------------------------------------------------------------------
\path[fill=bloblight]
  plot[smooth cycle,tension=0.8] coordinates {
    (5.65,1.30) (5.55,2.50) (6.05,3.80) (7.05,4.25)
    (8.75,4.15) (9.55,3.60) (9.80,2.10) (9.35,1.08)
    (8.35,0.72) (6.75,0.82)
  };

\path[fill=blobdark]
  plot[smooth cycle,tension=0.8] coordinates {
    (9.35,1.55) (9.55,2.55) (10.40,2.95) (11.45,2.75)
    (12.00,2.10) (11.80,1.15) (10.95,0.80) (9.85,0.95)
  };

\path[fill=blobdark,fill opacity=0.32,material]
  plot[smooth cycle,tension=0.55] coordinates {
    (6.7,0.95) (6.92,1.58) (6.1,2.02) (7.24,2.48)
    (6.35,3.18) (7.84,3.04) (8.32,3.46) (9.7,3.2) (8.62,2.92) (8.58,2.42)
    (8.9,1.94) (8.44,0.8) (8.05,1.7)
  };

% Converging streamlines
\draw[flow]
  plot[smooth,tension=0.65] coordinates {(7.65,2.52) (8.60,2.15) (9.55,1.8) (10.00,1.65)};
\draw[flow]
  plot[smooth,tension=0.65] coordinates {(8.85,3.25) (9.70,2.40) (10.10,2.15) (10.70,2.05)};
\draw[flow]
  plot[smooth,tension=0.65] coordinates {(8.4,2.95) (8.90,2.55) (9.60,2.13) (10.04,1.88)};
\draw[flow]
  plot[smooth,tension=0.65] coordinates {(7.55,1.95) (8.60,1.62) (10.5,1.55)};
\draw[flow]
  plot[smooth,tension=0.65] coordinates {(7.1,1.30) (8.50,1.0) (9.35,1.10) (10.92,1.50)};

\node[epslab] at (7.60,3.85) {$\varepsilon_{2\ell}(t-\tau_\ell)$};
\node[epslab] at (6.1,2.7) {$\widetilde{\mathcal{B}}_{2\ell}$};
\node[epslab] at (7.1,2) {$\widetilde{\mathcal{B}}_\ell$};
\node[epslab] at (10.7,2.55) {$\varepsilon_\ell(t)$};
\node[epslab] at (11.4,1.75) {$\mathcal{B}_\ell$};

\draw[dim] (5.65,0.28) -- (9.85,0.28);
\node[lab,below=2pt] at (7.68,0.32) {$2\ell$};
\draw[dim] (9.50,0.02) -- (11.70,0.02);
\node[lab,below=2pt] at (10.60,0.06) {$\ell$};

\node[font=\Large,above] at (9,-1.2){$\widetilde{\mathcal M}_\ell = \varepsilon_\ell(t)/\varepsilon_{2\ell}(t-\tau_\ell)\ \rightsquigarrow\ \mathcal M_\ell(t)$};
\node[font=\Large,below] at (8.40,-1.2) {{\bfseries (b)} {\sf Temporal inheritance}};

\draw[timearrow] (6.5,4.35) -- (11.5,4.35);
\node[font=\Large,above=2pt] at (6.5,4.35) {$t-\tau_\ell$};
\node[font=\Large,above=2pt] at (11.5,4.35) {$t$};

\end{tikzpicture}}
    \caption{Schematic representations of \textbf{(a)} the classical spatial-sampling bridge, $\tilde{\varepsilon}_{\tau_\ell}\sim\varepsilon_\ell$ for $\tau_\ell\sim\varepsilon_\ell^{-1/3}\ell^{2/3}$, and \textbf{(b)} dynamical inheritance. After the material doubling time $\tau_\ell$, the pre-image $\widetilde{\mathcal{B}}_\ell$ of $\mathcal{B}_\ell$ lies within the parent region $\widetilde{\mathcal{B}}_{2\ell}$. The resulting dynamical multiplier $\widetilde{\mathcal{M}}_\ell$ is statistically related to the Eulerian multiplier $\mathcal{M}_\ell$.}
    \label{fig:sketch_bridge}
\end{figure}
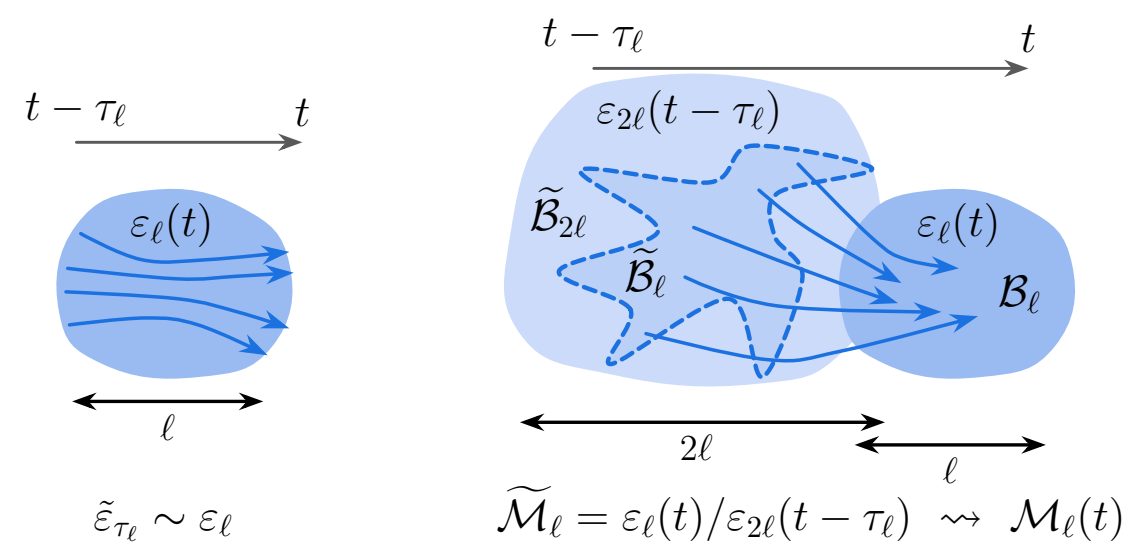

In the classical construction, illustrated in Fig.~\ref{fig:sketch_bridge}a, a spatial scale $\ell$ is first associated with its local eddy-turnover time. Using the refined-similarity estimate $\delta_\ell u^3/\ell\sim\varepsilon_\ell$ gives $\tau_\ell\sim\varepsilon_\ell^{-1/3}\ell^{2/3}$. The bridge then makes a second, stronger assumption: during this interval, a representative tracer samples the dissipation carried by the corresponding region sufficiently well that its temporal average can be identified with the instantaneous spatial average. These two steps lead to
\begin{equation}
    \tilde{\varepsilon}_{\tau_\ell}\sim\varepsilon_\ell,\qquad \tau_\ell=c\,\varepsilon_\ell^{-1/3}\ell^{2/3},
    \label{eq:bridge_classic}
\end{equation}
where the order-one constant $c$ accounts for the ambiguity in defining a turnover time. This is the spatial-sampling interpretation of the classical bridge.

Taken literally, Eq.~\eqref{eq:bridge_classic} identifies not only the characteristic scales of the two descriptions, but also their local dissipation fluctuations. This is difficult to reconcile with the results of the previous sections. Lagrangian time averages display approximately scale-stationary multipliers and the usual logarithmic cascade depth $\propto \log(\tau/\tau_{\rm c})$, whereas Eulerian spatial averages have scale-dependent multipliers and require a depth $\propto[\log(\ell/\ell_{\rm c})]^2$. 
This difference in scale organization makes a trajectory-wise interpretation of Eq.~\eqref{eq:bridge_classic} questionable. It does not, however, by itself imply that the two descriptions are statistically unrelated. We therefore test the stronger assumption of a one-to-one trajectory-wise correspondence directly.

\begin{figure}[h]
     \includegraphics[width=\linewidth]{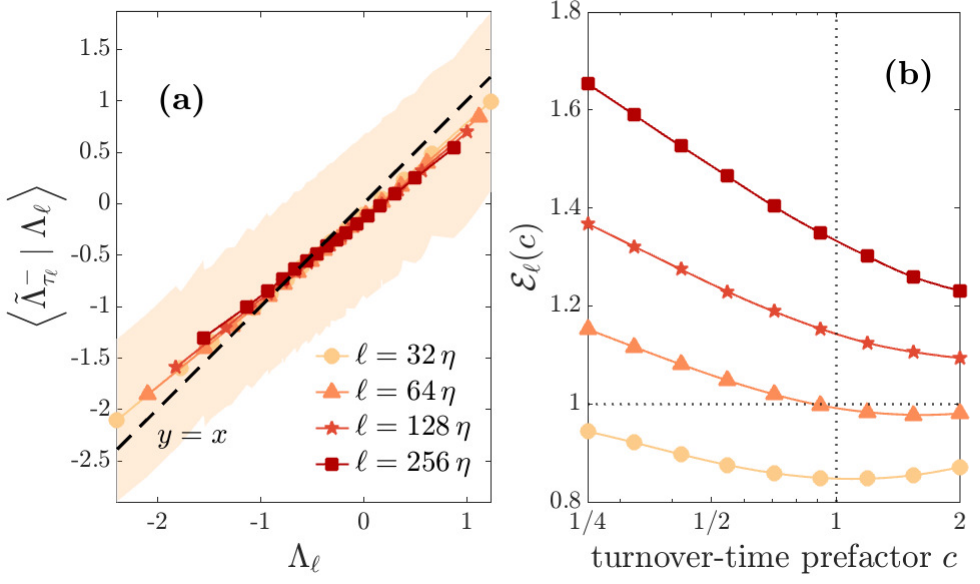}
     \caption{\textit{Direct test of the classical dissipation bridge at $R_\lambda=730$.} \textbf{(a)} Conditional average of the backward Lagrangian log-dissipation $\tilde{\Lambda}_{\tau_\ell}^- = \log(\tilde{\varepsilon}_{\tau_\ell}^-/\bar\varepsilon)$ given the Eulerian one $\Lambda_\ell = \log(\varepsilon_\ell/\bar\varepsilon)$, for $c=1$ and the indicated scales. The shaded area shows the conditional standard deviation, and the dashed line represents the direct bridge. \textbf{(b)} Normalized rms logarithmic difference between both quantities as a function of $c$. Dotted lines mark $c=1$ and a unit error.}
     \label{fig:direct_dissipation_bridge}
\end{figure}

For each tracer, we identify the Eulerian cell of size $\ell$ containing its position at the reference time $t$. The corresponding Eulerian dissipation $\varepsilon_\ell$ determines the local time $\tau_\ell = c\,\varepsilon_\ell^{-1/3}\ell^{2/3}$. It is compared with the backward Lagrangian average $\tilde{\varepsilon}_{\tau_\ell}^- =(1/\tau_\ell)\int_{t-\tau_\ell}^{t} \varepsilon(\boldsymbol{X}(s),s)\,\dd s$. Figure~\ref{fig:direct_dissipation_bridge}(a) conditions $\tilde{\Lambda}_{\tau_\ell}^- = \log(\tilde{\varepsilon}_{\tau_\ell}^-/\bar\varepsilon)$ on $\Lambda_\ell = \log(\varepsilon_\ell/\bar\varepsilon)$ for $c=1$, grouping cells into equal-population bins of the latter quantity. Figure~\ref{fig:direct_dissipation_bridge}(b) measures their normalized root-mean-square difference $\mathcal{E}_\ell(c)= \langle(\tilde{\Lambda}_{\tau_\ell}^--\Lambda_\ell)^2\rangle^{1/2} / [\operatorname{Var}(\Lambda_\ell)]^{1/2}$ as a function of the parameter $c$.

For $c=1$, the conditional averages in Fig.~\ref{fig:direct_dissipation_bridge}a increase with $\varepsilon_\ell$, confirming appreciable Eulerian--Lagrangian coupling. Nevertheless, the fluctuations among trajectories remain broad and the conditional means depart from the predicted diagonal. Correspondingly, $\mathcal{E}_\ell(c)$ remains of order unity, and no single choice of $c$ reduces it simultaneously at all scales. The classical bridge therefore fails as a one-to-one relation for a representative trajectory, rather than through a complete absence of coupling between the Eulerian and Lagrangian descriptions.

This failure has a simple physical interpretation. The Eulerian quantity $\varepsilon_\ell$ averages over the many fluid elements occupying a coarse-graining region $\mathcal B_\ell$ of size $\ell$ at time $t$, whereas $\tilde{\varepsilon}_{\tau_\ell}^-$ records the history of one such element. Their connection must therefore be considered at the level of the ensemble of trajectories ending in $\mathcal B_\ell$, rather than through a representative trajectory. The classical bridge replaces the resulting distribution by a representative trajectory, amounting to a mean-field approximation that Fig.~\ref{fig:direct_dissipation_bridge} rules out. Moreover, when this ensemble is followed backward in time, it is transported and deformed and can no longer be represented by the same spatial region. A refined bridge must consequently account for both trajectory-to-trajectory fluctuations and material deformation.

Once the one-to-one bridge is abandoned, the natural quantity to consider is the refinement between two successive spatial scales, measured by the Eulerian multiplier $\mathcal{M}_\ell=\varepsilon_\ell/\varepsilon_{2\ell}$. A conventional spatial cascade would associate this refinement with an approximately scale-independent multiplier. Section~\ref{sec:eulerian} showed that this is not the case. Its scale dependence need not, however, be devoid of structure and we next examine how multiplier statistics themselves evolve across scales.

\begin{figure*}[t]
    \includegraphics[width=\linewidth]{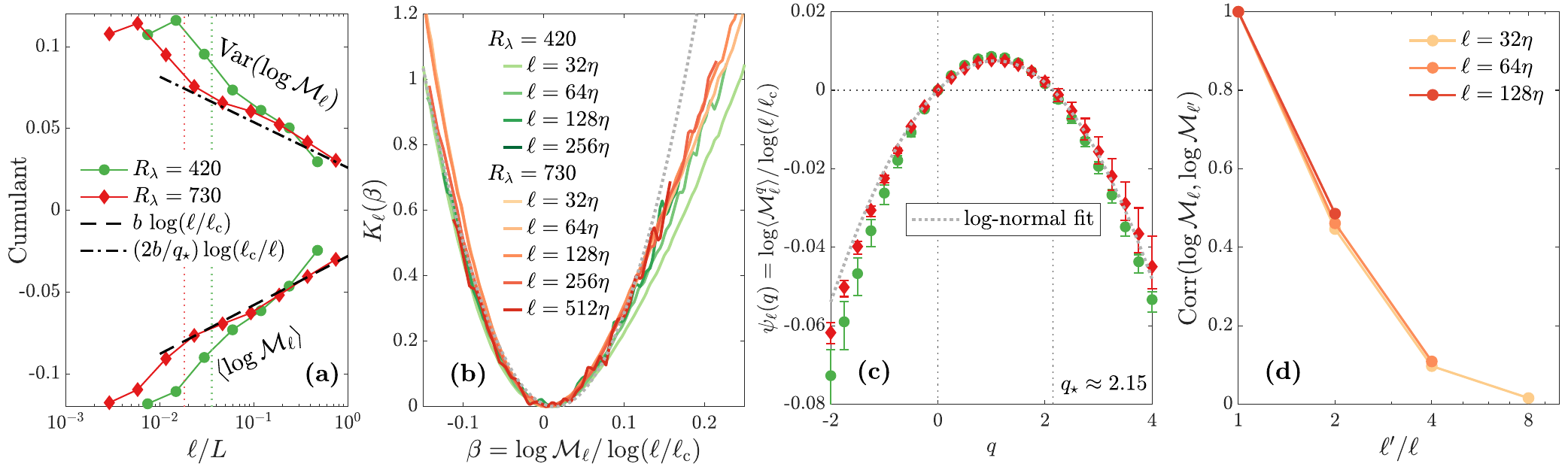}
    \caption{\textit{Multifractal statistics of Eulerian dissipation multipliers.} \textbf{(a)} Mean and variance of $\mathcal{X}_\ell = \log \mathcal{M}_\ell = \log(\varepsilon_\ell/\varepsilon_{2\ell})$ at $R_\lambda=420$ and $730$. Lines show the logarithmic fits indicated in the legend with $b=0.013$ and $q_\star\simeq2.15$; vertical dotted lines mark the Taylor microscales. \textbf{(b)} Finite-scale rate functions of $\beta=\mathcal{X}_\ell/\log(\ell/\ell_{\rm c})$ at the indicated scales. The dotted curve is the parabolic (log-normal) fit $K(\beta)=(\beta-bq_\star/2)^2/(2b)$, with the same values of $b$ and $q_\star$ as in panel (a). \textbf{(c)}~Multiplier moment exponents $\psi_\ell(q)=\log\langle\mathcal M_\ell^q\rangle/\log(\ell/\ell_{\rm c})$; error bars indicate their scale-to-scale variation. The dotted curve is $\psi(q)=(b/2)q(q_\star-q)$. \textbf{(d)}~Correlation coefficients $\operatorname{Corr}(\mathcal{X}_\ell,\mathcal{X}_{\ell'})$ between the logarithms of concentric Eulerian multipliers at $R_\lambda=730$, plotted against the scale ratio $\ell'/\ell$ for the indicated reference scales.}
    \label{fig:eulerian_multipliers} 
\end{figure*}

Figure~\ref{fig:eulerian_multipliers} establishes one of the central results of this work: the Eulerian multipliers possess a multifractal scale organization of their own. Their lack of scale invariance is not an arbitrary drift of their distributions. The mean and variance of $\mathcal {X}_\ell=\log\mathcal{M}_\ell$ vary linearly with the logarithmic distance from the cutoff scale, with compatible behaviors at both Reynolds numbers (Fig.~\ref{fig:eulerian_multipliers}a). Correspondingly, the rescaled variable $\beta=\mathcal{X}_\ell/\log(\ell/\ell_{\rm c})$ obeys a finite-scale large-deviation principle, while the multiplier moments follow power laws,
\begin{equation}
p_\ell(\beta)\asymp(\ell/\ell_{\rm c})^{K(\beta)}, \qquad \left\langle\mathcal M_\ell^q\right\rangle \sim (\ell/\ell_{\rm c})^{\psi(q)}.
\end{equation}
These scaling forms are understood over the resolved scale interval. Indeed, nested averaging regions imply $0<\mathcal{M}_\ell \leq 8$, so positive-order moments remain bounded and branches with $\psi(q)<0$ cannot persist as $\ell/\ell_{\rm c}\to0$. Within this finite-range interpretation, the collapse of the rate functions in  Fig.~\ref{fig:eulerian_multipliers}b and the nonlinear, approximately scale-independent function $\psi(q)$ in Fig.~\ref{fig:eulerian_multipliers}c provide equivalent representations of this multifractality. The conventional logarithmic cascade depth that is absent from the coarse-grained Eulerian dissipation therefore reappears one level lower in the hierarchy, in the increments $\log\varepsilon_\ell-\log\varepsilon_{2\ell}$.

The collapsed rate functions and moment exponents further show that this multiplier multifractality is approximately, but not exactly, log-normal. Over the central range of fluctuations, $K(\beta)$ is close to the parabolic form shown in Fig.~\ref{fig:eulerian_multipliers}b, while $\psi(q)$ is reasonably described by the corresponding quadratic approximation in Fig.~\ref{fig:eulerian_multipliers}c. Equivalently, $\mathcal{X}_\ell$ is approximately Gaussian, although systematic departures in the tails show that the log-normal description is not complete. A notable feature of the quadratic approximation is its nontrivial zero at $q=q_\star\simeq2.15$: $\psi(q_\star)=0$, $\langle\mathcal{M}_\ell^{q_\star}\rangle
\simeq \text{const}$. Within the log-normal approximation, the same condition follows from the scale-independent relation $\langle\mathcal{X}_\ell\rangle\simeq-(q_\star/2)\operatorname{Var}(\mathcal{X}_\ell)$ observed in Fig.~\ref{fig:eulerian_multipliers}a. The positive root $q_\star$, unlike the trivial normalization condition at $q=0$, expresses a balance between the typical refinements with $\mathcal M_\ell<1$ and the rarer intense events with $\mathcal M_\ell>1$. If it persists over a broader inertial range, this approximately conserved moment would constitute a generalized constraint on the spatial refinement process, and hence on any dynamical bridge relating it to Lagrangian multipliers.

The observed intermittency of multipliers differs fundamentally from a conventional spatial cascade where a dyadic refinement would have the same statistics at every level. Here, instead, the fluctuations of a single refinement become increasingly broad as $\ell$ moves away from $\ell_{\rm c}$: the multiplier retains a memory of how far the cascade has progressed. The equal-time spatial refinement is therefore not an autonomous, scale-local step. 
The scale dependence of the Eulerian multiplier is instead a constraint that any connection between spatial and temporal descriptions must account for. A possible interpretation is that it reflects the material histories accumulated before fluid elements are assembled into the child region. Transport, deformation, and mixing provide natural mechanisms for such a reorganization, but the present measurements do not directly establish that they transform the scale-stationary temporal multipliers into the observed spatial multipliers. The measured multiplier multifractality therefore provides a quantitative constraint on, rather than a direct demonstration of, such a dynamical bridge.

The multifractality of Eulerian dissipation multipliers provides a statistical interpretation of the squared-logarithmic large-deviation depth found in Sec.~\ref{sec:eulerian}. Along a dyadic hierarchy $\ell_j = \ell_{\rm c} / 2^j$ with $\ell_n=\ell$, the identity $\Lambda_\ell=\sum_{j=1}^n \mathcal{X}_{\ell_j}$ gives $\left\langle\Lambda_{\ell}\right\rangle = \textstyle \sum_{j=1}^n \left\langle \mathcal{X}_{\ell_j}\right\rangle$ and $\operatorname{Var}(\Lambda_{\ell}) = \textstyle \sum_{j=1}^n\operatorname{Var}(\mathcal{X}_{\ell_j}) +2\sum_{i<j}\operatorname{Cov}(\mathcal{X}_{\ell_i},\mathcal{X}_{\ell_j})$. The number of multiplier increments grows as $n\sim |\log(\ell/\ell_{\rm c})|$, while the mean and variance of each increment grow linearly with its logarithmic distance from $\ell_{\rm c}$. Summing these marginal contributions therefore produces a quadratic dependence $\sim [\log(\ell/\ell_{\rm c})]^2$. For the variance, however, the measured covariance contributions are comparable to the sum of marginal variances and cannot be neglected. Figure~\ref{fig:eulerian_multipliers}d shows that the correlation coefficients depend primarily on the scale ratio and decrease rapidly over the available separations. If this behavior extends to a summable correlation kernel in logarithmic scale, the accumulated interscale covariances can also contribute at order $n^2$, modifying the coefficient while preserving the squared-logarithmic dependence. The measurements support this interpretation over the resolved hierarchy, although the limited range of scale pairs above the Taylor microscale precludes establishing its asymptotic validity.

The Taylor microscale $\lambda$ approximately marks the lower boundary of this multiplier regime. In Fig.~\ref{fig:eulerian_multipliers}a, the logarithmic behavior of the multiplier cumulants is observed mainly for $\ell\gtrsim\lambda$, while smaller scales display a Reynolds-number-dependent departure. Consistently, the rate functions in Fig.~\ref{fig:eulerian_multipliers}b are approximately compatible from $64\eta$ upward, whereas the result at $32\eta$, below or close to $\lambda$, remains distinct. The relevance of $\lambda$ has a natural space-time interpretation: the Taylor microscale is, up to a conventional factor, the distance over which dissipative structures are swept during one Kolmogorov time. The observed crossover is therefore consistent with a transition from nearly frozen spatial sampling at smaller scales to a regime in which temporal evolution becomes important. Uniform sweeping alone, however, cannot alter equal-time multiplier statistics. Their nontrivial scale dependence must involve the evolution and deformation of the dissipation field during transport. This motivates the material-inheritance construction illustrated in Fig.~\ref{fig:sketch_bridge}b, which provides a way to make explicit the distinction between an instantaneous Eulerian parent and the material history contributing to the present child.

The fluid elements occupying $\mathcal{B}_\ell$ at time $t$ carry dissipation fluctuations acquired during their previous evolution. Denote by $\widetilde{\mathcal{B}}_\ell(t-\tau)$ the pre-image of $\mathcal{B}_\ell(t)$ under the flow. As $\tau$ increases, this material region is transported, stretched, and folded. We define $\tau_\ell$ as the first backward time at which the characteristic size of the pre-image reaches $2\ell$, and denote by $\widetilde{\mathcal B}_{2\ell}(t-\tau_\ell)$ a region of size $2\ell$ containing this pre-image. We henceforth denote by $\varepsilon_\ell(t)$ the dissipation averaged over the child region $\mathcal{B}_\ell(t)$, and by $\varepsilon_{2\ell}(t-\tau_\ell)$ that averaged over its dynamically selected parent $\widetilde{\mathcal{B}}_{2\ell}(t-\tau_\ell)$ (see Fig.~\ref{fig:sketch_bridge}b). With this convention, the dynamical refinement reads
\begin{equation}
    \widetilde{\mathcal{M}}_\ell(t) = \frac{\varepsilon_\ell(t)} {\varepsilon_{2\ell}(t-\tau_\ell)}.
    \label{eq:dynamical_multiplier}
\end{equation}
Unlike the Eulerian multiplier $\mathcal{M}_\ell(t)$, whose parent and child are sampled simultaneously, $\widetilde{\mathcal{M}}_\ell$ compares the present child with its dynamically selected parent in the past. It therefore represents temporal inheritance rather than instantaneous spatial refinement.

To connect this dynamical refinement to the Lagrangian cascade, we use the correspondence between space and time only at the level of filtering. For an inertial-range scale $\ell$, averaging the instantaneous dissipation over all tracers in a region of size $\ell$ suppresses fluctuations associated with times shorter than the corresponding turnover time $\tau_\ell$. 
We therefore use the following statistical representation: after averaging over the tracer ensemble occupying a spatial region, the instantaneous spatial average can be related to time-coarse-grained dissipation along the contributing trajectories. For uniformly distributed tracers, we further hypothesize the following approximate statistical correspondences:
\begin{align}
    \varepsilon_\ell(t) & \overset{\rm stat}{\sim} \frac{1}{N_\ell}  \sum_{\boldsymbol X_i(t)\in\mathcal B_\ell(t)} \tilde{\varepsilon}_{\tau_\ell,i}(t), \nonumber \\
    \varepsilon_{2\ell}(t-\tau_\ell) &\overset{\rm stat}{\sim} \frac{1}{N_{2\ell}} \sum_{\boldsymbol X_j(t-\tau_\ell)\in\widetilde{\mathcal B}_{2\ell}(t-\tau_\ell)} \tilde{\varepsilon}_{\tau_{2\ell},j}(t-\tau_\ell), \label{eq:spatiotemporal_filtering}
\end{align}
where $N_\ell$ and $N_{2\ell}$ are the numbers of tracers in the child region $\mathcal B_\ell(t)$ and in the enclosing parent region $\widetilde{\mathcal B}_{2\ell}(t-\tau\ell)$, respectively. The latter contains the material pre-image of the child as well as additional fluid elements, so the two tracer counts generally differ. The proposed statistical correspondences concern averages over these ensembles and do not assert that the temporal average along an individual tracer equals the corresponding spatial average.

Along each trajectory ending in the child region, the two temporal averages are related by the scale-adapted Lagrangian multiplier
\begin{equation}
    \mathcal{N}_{\ell,i}(t-\tau_\ell) = \frac{\tilde{\varepsilon}_{\tau_\ell,i}(t-\tau_\ell)}{\tilde{\varepsilon}_{\tau_{2\ell},i}(t-\tau_\ell)}.
    \label{eq:scale_adapted_lag_multiplier}
\end{equation}
This is the same temporal refinement as $\mathcal N_\tau$ in Eq.~\eqref{eq:def_lag_multipliers}, except that the two averaging times are now selected by the spatial scales $\ell$ and $2\ell$; it reduces exactly to $\mathcal N_{\tau_\ell}$ when $\tau_{2\ell}=2\tau_\ell$. Equations~\eqref{eq:dynamical_multiplier} and \eqref{eq:spatiotemporal_filtering} then give
\begin{equation}
    \widetilde{\mathcal M}_\ell(t) \overset{\rm stat}{\sim} \frac{N_{2\ell}}{N_\ell} \frac{\sum_{\boldsymbol X_i(t)\in\mathcal B_\ell(t)}\mathcal N_{\ell,i}\,\tilde{\varepsilon}_{\tau_{2\ell},i}}{\sum_{\boldsymbol X_j(t-\tau_\ell)\in\widetilde{\mathcal B}_{2\ell}(t-\tau_\ell)}\tilde{\varepsilon}_{\tau_{2\ell},j}},
    \label{eq:dynamical_lagrangian_connection}
\end{equation}
where the common time argument $t-\tau_\ell$ has been omitted on the right-hand side. Thus, $\widetilde{\mathcal M}_\ell$ is not identified with a single Lagrangian multiplier. 
Rather, it involves the temporal refinements collectively carried by the fluid elements 
with their relative weights set by the parent-scale dissipation. Material deformation and mixing determine which histories contribute to the present spatial region and how they are assembled.

The final step is to relate this dynamical refinement to the Eulerian multiplier. Both quantities contain the same present child but refer to different parents:
\begin{equation}
    \mathcal M_\ell(t) = \widetilde{\mathcal M}_\ell(t)\,\frac{\varepsilon_{2\ell}(t-\tau_\ell)}{\varepsilon_{2\ell}(t)}.
    \label{eq:dynamical_eulerian_connection}
\end{equation}
Here $\varepsilon_{2\ell}(t)$ denotes the dissipation in the instantaneous concentric parent $\mathcal B_{2\ell}(t)$, whereas $\varepsilon_{2\ell}(t-\tau_\ell)$ refers to the dynamically selected parent defined above. The second factor measures the difference between the dynamically selected material parent and the instantaneous Eulerian parent. It therefore contains the effects of transport, deformation, mixing, and temporal evolution of the dissipation field. Because this factor fluctuates and may correlate with $\widetilde{\mathcal M}_\ell$, the dynamical and Eulerian multipliers need not coincide. Equation~\eqref{eq:dynamical_eulerian_connection} consequently provides a statistical decomposition of the Eulerian refinement into a material contribution and a parent-evolution contribution. It does not require, and does not establish, a one-to-one correspondence between Eulerian and Lagrangian multipliers.

The decomposition~\eqref{eq:dynamical_eulerian_connection} turns the material-inheritance picture into a
falsifiable statement. If the scale dependence of $\mathcal{M}_\ell$ arises
from the assembly of material histories, the dynamical multiplier
$\widetilde{\mathcal{M}}_\ell$ should be approximately scale-stationary, with
statistics close to those of the temporal multipliers $\mathcal{N}_{\tau_\ell}$,
while the logarithmic drift of Fig.~9(a) should be carried by the
parent-evolution factor $\varepsilon_{2\ell}(t-\tau_\ell)/\varepsilon_{2\ell}(t)$
and its correlation with $\widetilde{\mathcal{M}}_\ell$. Conversely, a
dynamical multiplier as scale-dependent as $\mathcal{M}_\ell$ would show that
the Eulerian refinement is not explained by material transport. Measuring
$\widetilde{\mathcal{M}}_\ell$ requires following the pre-images of
coarse-graining regions with tracer densities beyond those used here.

%\textcolor{blue}{%
%The resulting picture is therefore a refinement of the classical bridge rather than another one-to-one bridge relation. The trajectory-wise identification in Eq.~\eqref{eq:bridge_classic} is not supported by the data, while the Eulerian and Lagrangian quantities remain statistically coupled. The Lagrangian description exposes the scale organization of dissipation histories along trajectories, whereas the Eulerian description captures the organization of those histories after they have been spatially assembled and deformed by the flow. The material-inheritance construction makes this distinction explicit and provides a framework in which the two different multiplier organizations can coexist. Establishing whether material deformation and mixing quantitatively generate the observed Eulerian multiplier statistics will require a direct analysis of the associated material pre-images and histories.
%}

%%%%%%%%%%%%%%%%%%%%%%%%%%%%%%%%%%%%%%	

\section{Conclusions}
\label{sec:conclusion}

These results show that the intermittency of turbulent dissipation has
distinct statistical organizations under spatial and temporal
coarse-graining. A central organizing idea in turbulence theory is that
intermittency arises from a cascade process that can be expressed through
multifractal statistics of energy dissipation. Our results show that this
picture cannot be represented by a single, frame-independent cascade
structure. Lagrangian time averages display an approximately
scale-stationary temporal cascade, whereas Eulerian coarse-grained
dissipation exhibits a distinct spatial organization with scale-dependent
multipliers and a squared-logarithmic cascade depth. The commonly assumed
equivalence between Eulerian and Lagrangian dissipation statistics,
mediated by bridge relations, therefore cannot be interpreted as a
trajectory-wise identification. Rather, the results indicate that the
temporal organization revealed along Lagrangian trajectories provides an
important description of dissipation intermittency, while its Eulerian
manifestation reflects a different spatial organization.

To reach this conclusion, we performed a systematic comparison of
Eulerian and Lagrangian statistics of energy dissipation in homogeneous
isotropic turbulence. By combining analyses of single-point
distributions, correlation functions, coarse-grained fields,
large-deviation properties, and multiplier statistics, we showed that
the statistical organization of dissipation depends fundamentally on
the manner in which it is sampled and cannot be captured by a single
universal cascade description.

In the Eulerian frame, the spatial correlations and coarse-grained properties
of the instantaneous dissipation field do not support the existence of a
conventional scale-stationary multiplicative cascade. In particular, the
absence of logarithmic spatial correlations, the nonlinear dependence of
cumulants on scale, and the pronounced scale dependence of Eulerian multipliers
rule out a conventional log-divisible cascade interpretation of the
spatially coarse-grained Eulerian dissipation in the inertial range.  At the same time,
this measure still shows a large deviation form with squared-logarithmic depth.
Furthermore, the Eulerian multipliers possess a nontrivial multifractal
organization. This surprising, yet self-consistent picture, forces us to
reimagine the physically appealing, but certainly incomplete, cascade
phenomenology and go beyond the scale-stationary processes which underlies much
of conventional turbulence theory.

In contrast, dissipation sampled along Lagrangian trajectories displays the
defining signatures of a multiplicative cascade in time. These include
logarithmic temporal correlations, linear scaling of the mean and variance of
the logarithm of the coarse-grained dissipation with the logarithm of the time
scale, and approximately scale-invariant statistics of Lagrangian multipliers.
The temporally coarse-grained Lagrangian dissipation is not log-normal but
still follows a well-defined large-deviation principle. Thus  the Lagrangian
dissipation is consistent with a log-divisible stochastic process organized by
a temporal multiplicative cascade.

Building on these findings, we revisited classical bridge relations that
aim to connect Eulerian and Lagrangian dissipation statistics through
dimensional arguments. We showed that such relations implicitly rely on
a mean-field assumption, namely that dissipation fluctuations experienced
by different Lagrangian trajectories within the same spatial region are
effectively uniform. Our results demonstrate that this assumption is
violated for dissipation, leading to a breakdown of the classical bridge
relation as a trajectory-wise identification. This breakdown does not
imply an absence of Eulerian--Lagrangian coupling: the two descriptions
remain appreciably correlated, but their fluctuations cannot be
represented by a single trajectory or by a single choice of
turnover-time prefactor.

We therefore formulated a material-inheritance framework in which the
present Eulerian coarse-graining region is related to the material
histories of the fluid elements contributing to it. The corresponding
dynamical refinement compares the present child region with a dynamically
selected material parent and can be expressed statistically in terms of
the temporal refinements carried by the trajectories that contribute to
the child. The Eulerian multiplier then contains, in addition, a factor
describing the evolution between this dynamically selected parent and the
instantaneous Eulerian parent. This construction provides a statistical
framework for connecting spatial and temporal descriptions without
requiring a one-to-one mapping between Eulerian and Lagrangian
multipliers.

More broadly, these results indicate that the distinction between
Eulerian and Lagrangian intermittency is not simply a distinction between
two measurements of the same cascade. The Lagrangian description exposes
the scale organization of dissipation histories along material
trajectories, while the Eulerian description captures the organization
that results when these histories are spatially assembled, transported,
and deformed by the flow. The present analysis does not establish a
direct causal transformation from the temporal cascade into the observed
Eulerian multiplier statistics. Instead, it identifies material
inheritance as a possible framework for relating the two descriptions
and shows that any such connection must be consistent with the distinct
multiplier statistics and cascade depths observed in the two sampling
procedures.

Future work may test this framework more directly by following material
pre-images of Eulerian coarse-graining regions and examining how their
dissipation histories are transported, deformed, and assembled across scales.
While it may be tempting to connect our results on dissipation to the velocity
increment statistics directly --- since $\epsilon_\ell$ is treated as a proxy
for the energy flux $\sim \delta_\ell v^3/\ell$ --- the absence of a clear
multifractal statistics in the former does not preclude multifractality in the
latter.  In any case, the present results suggest that models of intermittency
may benefit from distinguishing the temporal organization of fluctuation
histories from the spatial statistics that emerge when those histories are
sampled and assembled by the flow. A particularly well-adapted framework to 
factor this in could be quasi-Lagrangian approaches and models~\cite{Belinicher1987,boffetta2002lagrangian,MitraPandit2004} and its generalisations~\cite{RayQL} 
including those involving dynamical time-rescaling as suggested by hidden symmetry 
arguments~\cite{mailybaev2022hidden}.

\begin{acknowledgments}
This research has been partly supported by the Agence Nationale de la Recherche (Grant No.\ ANR-25-CE30-4603-01). RM acknowledges the Infosys-TIFR Leading Edge Travel Grant 2025 which facilated part of this research.  SM acknowledges the IITK Initiation Grant project IITK/ME/2024316 and support from the Govt. of India Grant ANRF/ECRG/2024/002467/ENS.  JB and SSR acknowledges  the Indo–French Centre for the Promotion of Advanced Scientific Research (IFCPAR/CEFIPRA, project no. 6704-1) for support.  The simulations were performed on the ICTS clusters Mario, Tetris, and Contra, using OPAL infrastructure from Université Côte d’Azur, and HPC resources from GENCI-TGCC (Grant t2016-2as027). RM and SSR acknowledge the support of the Department of Atomic Energy, Government of India, under project no. RTI4019.
\end{acknowledgments}
	
\bibliography{references.bib}
	
\end{document}